\documentclass{article}

\usepackage{moreverb}
\usepackage[colorlinks,bookmarksopen,bookmarksnumbered,citecolor=red,urlcolor=red]{hyperref}
\usepackage{xspace}
\usepackage{listings}
\usepackage{dcolumn}
\usepackage{tabularx}

\usepackage{times}

\usepackage{tikz}
\usetikzlibrary{arrows,automata,shapes.symbols}
\usetikzlibrary{external}
\tikzset{external/only named=true}

\newcommand{\nameSys}{C2Eif\xspace}
\newcommand{\titledpar}[1]{\vskip 3pt \textbf{#1}}
\newcommand{\transl}{\ensuremath{\mathcal{T}}}
\newcommand{\tr}[1]{\mathcal{T}_{\mathrm{#1}}}

\newcommand{\libname}[1]{\hbox{\tt #1}}
\newcommand{\headname}[1]{\hbox{\tt #1}}
\newcommand{\progname}[1]{\hbox{\tt #1}}
\newcommand{\cc}[1]{\mbox{\lstinline[language=CustomC]|#1|}}
\newcommand{\eif}[1]{\mbox{\lstinline[language=OOSC2Eiffel]|#1|}}

\newenvironment{codesnip}{%
\begin{normalsize}}{%
\end{normalsize}}

\usepackage{fmtcount}

\usepackage{amsmath,booktabs,rotating}
\usepackage{amsthm}
\theoremstyle{plain}
\theoremstyle{definition}
\newtheorem{example}{Example}

\begin{document}

\title{C to O-O Translation: Beyond the Easy Stuff\footnote{This work was partially supported by ETH grant ``Object-oriented reengineering environment''.}}

\author{
  Marco~Trudel
  $\cdot$
  Carlo A.~Furia
  $\cdot$
  Martin~Nordio \\
  Bertrand~Meyer
  $\cdot$
  Manuel~Oriol
}

\date{19 April 2013}

\maketitle

\begin{abstract}
Can we reuse some of the huge code-base developed in C to take advantage of modern programming language features such as type safety, object-orientation, and contracts?
This paper presents a source-to-source translation of C code into Eiffel, a modern object-oriented programming language, and the supporting tool \nameSys.
The translation is completely automatic and handles the entire C language (ANSI, as well as many GNU C Compiler extensions, through CIL) as used in practice, including its usage of native system libraries and inlined assembly code.
Our experiments show that \nameSys can handle C applications and libraries of significant size (such as \libname{vim} and \libname{libgsl}), as well as challenging benchmarks such as the GCC torture tests.
The produced Eiffel code is functionally equivalent to the original C code,
and takes advantage of some of Eiffel's features to produce safe and easy-to-debug translations.


\end{abstract}

\maketitle

\section{Introduction}\label{sec:intro}

Programming languages have significantly evolved since the original design of C in the 1970's as a ``system implementation language''~\cite{C-history} for the Unix operating system.
C was a high-level language by the standards of the time, but it is pronouncedly low-level compared with modern programming paradigms, as it lacks advanced features---static type safety, encapsulation, inheritance, and contracts~\cite{OOSC2}, to mention just a few---that can have a major impact on programmer's productivity and on software quality and maintainability.

C still fares as the most popular general-purpose programming language~\cite{popularity}, and countless C applications are being actively written and maintained, that take advantage of the language's conciseness, speed, ubiquitous support, and huge code-base.
An automated solution to translate and integrate C code into a modern language would combine the large availability of C programs in disparate domains with the integration in a modern language that facilitates writing safe, robust, and easy-to-maintain applications.

The present paper describes the fully automatic translation of C programs into Eiffel, an object-oriented programming language, and the tool \nameSys, which implements the translation.
While the most common approaches to re-use C code in other host languages are based on ``foreign function APIs'' (see Section~\ref{sec:rw} for examples), source-to-source translation solves a different problem, and has some distinctive benefits: the translated code can take advantage of the high-level nature of the target language and of its safer runtime.

\titledpar{Main features of \nameSys.}
Translating C to a high-level object-oriented language is challenging because it requires adapting to a more abstract memory representation, a tighter type system, and a sophisticated runtime that is not directly accessible.  There have been previous attempts to translate C into an object-oriented language (see the review in Section~\ref{sec:rw}).
A limitation of the resulting tools is that they hardly handle the trickier or specialized parts of the C language~\cite{EllisonRosu12}, which it is tempting to dismiss as unimportant ``corner cases'', but figure prominently in real-world programs; examples include calls to pre-compiled C libraries (e.g., for I/O), inlined assembly, and unrestricted branch instructions including \cc{setjmp} and \cc{longjmp}.

One of the distinctive features of the present work is that it does not stop at the core features but extends over the often difficult ``last mile'': it covers the entire C language as used in practice.
The completeness of the translation scheme is attested by the set of example programs to which the translation was successfully applied, as described in Section~\ref{sec:eval}, including major applications such as the \progname{vim} editor (276 KLOC), major libraries such as \progname{libgsl} (238 KLOC), and the challenging ``torture'' tests for the GCC C compiler.

\nameSys is available at \url{http://se.inf.ethz.ch/research/c2eif}. The webpage includes \nameSys's sources, pre-compiled binaries, source and binaries of all translated programs of Table~\ref{tab:eval}, and a user guide.

Sections~\ref{sec:principles}--\ref{sec:impl} describe the distinctive features of the translation:
it supports the complete C language (including pointer arithmetic, unrestricted branch instructions, and function pointers) with its native system libraries;
it complies with ANSI C as well as many GNU C Compiler extensions through the CIL framework~\cite{NeculaMcPeakRahulWeimer};
it is fully automatic, and it handles complete applications and libraries of significant size;
the generated Eiffel code is functionally equivalent to the original C code (as demonstrated by running thorough test suites), and takes advantage of some advanced features, such as strong typing and contracts, to facilitate debugging some programming mistakes.

\begin{center}
\framebox{\parbox{0.9\textwidth}{\centering \emph{In our experiments, \nameSys translated completely automatically over 900,000 lines of C code from real-world applications, libraries, and testsuites, producing functionally equivalent$\footnotemark$ Eiffel code.}}}
\end{center}
\footnotetext{As per standard regression testsuites and general usage.}

\titledpar{Safer code.}
Translating C code to Eiffel with \nameSys is quite useful to reuse C programs in a modern environment, and is the main motivation behind this work.
However, it also implies other valuable side-benefits, which we demonstrate in Section~\ref{sec:eval}.
First, the translated code blends reasonably well with hand-written Eiffel code because it is not a mere transliteration from C; it is thus modifiable with Eiffel's native tools and environments (EiffelStudio and related analysis and verification tools).
Second, the translation automatically introduces simple contracts, which help detect recurring mistakes such as out-of-bound array access or null-pointer dereferencing.
To demonstrate this, Section~\ref{sec:bugs} discusses how we easily discovered a few unknown bugs in widely used C programs (such as \libname{libgmp}) just by translating them into Eiffel and running standard tests.
While the purpose of \nameSys is not to debug C programs, the source of errors is usually more evident when executing programs translated into Eiffel---either because a contract violation occurs, or because the Eiffel program fails sooner, before the effects of the error propagate to unrelated portions of the code.
The translated C code also benefits from the tighter Eiffel runtime, so that certain buffer overflow errors are harder to exploit than in native C environments.

\titledpar{Why Eiffel?}
We chose Eiffel as the target language not only out of our familiarity with it, but also because it offers features that complement C's, such as an emphasis on correctness~\cite{TFNM11-SEFM11} through the native support of contracts.

Another reason for the choice has to do with the size of Eiffel's community of developers, which is quite small compared to those of other object-oriented languages such as Java or C\#.
While Java users can avail of countless libraries and frameworks offering high-quality implementations of a wide array of functionalities, Eiffel users often have to implement such functionalities themselves from scratch.
An automatic translation tool could therefore bring immediate benefits to reuse the rich C code-base in Eiffel; in fact, translations produced by \nameSys are already being used by the Eiffel community.

Finally, Eiffel uncompromisingly epitomizes the object-oriented paradigm; hence translating C into it cannot take the shortcut of merely transliterating similar constructs (as it would have been possible, for example, with C++).
The results of the paper are thus applicable to other object-oriented languages.\footnote{As proof of this, the first author has made some progress towards applying the technique described in this paper to automatic translations from C to Java; and has recently been granted startup funding to support this development and turn it into a commercial product.}

\titledpar{}
An abridged version of this paper was presented at WCRE 2012~\cite{TFNMO12-WCRE12}; a companion paper presents the tool \nameSys from the user's point of view~\cite{C2Eif-demo}.


\section{Overview and Architecture}\label{sec:principles}

\nameSys is a compiler with graphical user interface that translates C programs to Eiffel programs.
The translation is a complete Eiffel program which replicates the functionality of the C source program.
\nameSys is implemented in Eiffel.
\begin{figure}[!htb]
\centering
  \includegraphics[width=1\columnwidth]{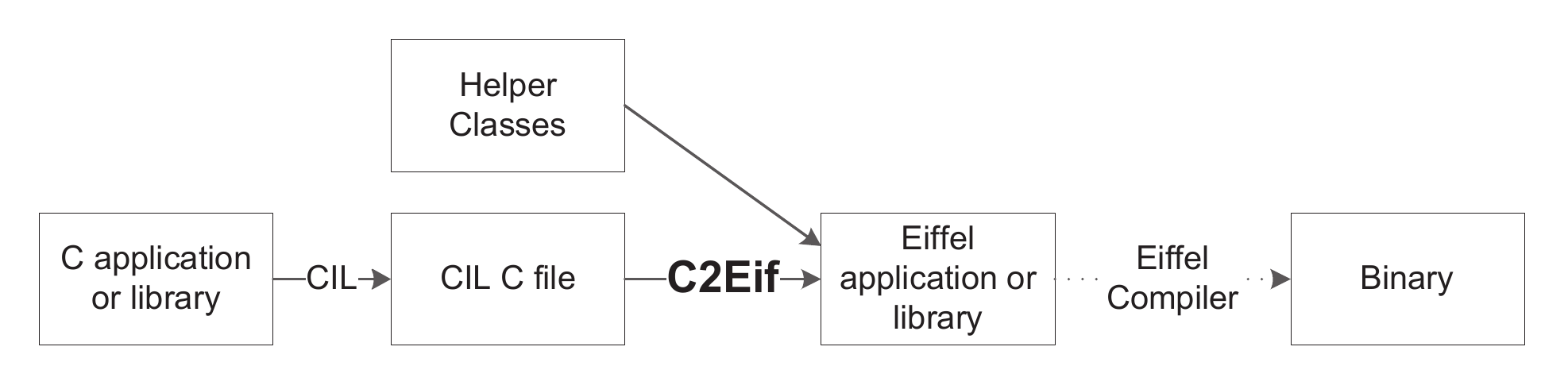}
  \caption{Usage workflow of \nameSys.}\label{fig:comp}
\end{figure}

\titledpar{High-level view.}  Figure~\ref{fig:comp} shows the overall
picture of how \nameSys works.  \nameSys inputs C projects
(applications or libraries) processed with the C Intermediate Language
(CIL) framework.  CIL~\cite{NeculaMcPeakRahulWeimer} is a C front-end that simplifies
programs written in ANSI C or using the GNU C Compiler extensions into
a restricted subset of C amenable to program transformations; for
example, there is only one form of loop in CIL.  Using CIL input to
\nameSys ensures complete support of the whole set of C statements, without
having to deal with each of them explicitly (although it also introduces some limitations, discussed in Section~\ref{sec:limitations}).
\nameSys then translates CIL
programs to Eiffel projects consisting of collections of classes
that rely on a small set of Eiffel helper classes
(described below). Such projects can be
compiled with any standard Eiffel compiler.

\titledpar{Incremental translation.}
\nameSys implements a translation $\transl$ from CIL C to Eiffel as a series $T_1, \ldots, T_n$ of successive incremental transformations on the Abstract Syntax Tree.
Every transformation $T_i$ targets exactly one language aspect (for example, loops or inlined assembly code) and produces a program in an intermediate language $L_i$ which is a mixture of C and Eiffel constructs:
the code progressively \textit{morphs} from C to Eiffel code.
The current implementation uses around 45 such transformations (i.e., $n = 45$).
Combining several simple transformations improves the decoupling among different language constructs and facilitates reuse (e.g., to implement a translator of C to Java) and debugging:
the intermediate programs are easily understandable by programmers familiar with both C and Eiffel.

\titledpar{Helper classes.}  The core of the translation from C to
Eiffel must ensure that Eiffel applications have access to objects
with the same capabilities as their counterparts in C; for
example, an Eiffel class that translates a C \cc{struct} has to
support field access and every other operation defined on
\cc{struct}s.  Conversely, C \emph{external} pre-compiled code may
also have to access the Eiffel representations of C constructs; for
example, the Eiffel translation of a C program calling \cc{printf} to
print a local string variable \cc{str} of type \cc{char*} must grant
\cc{printf} access to the Eiffel object that translates \cc{str}, in
conformance with C's conventions on strings.  To meet these
requirements, \nameSys includes a limited number of hand-written
helper Eiffel classes that bridge the Eiffel and C environments; their
names are prefixed by \eif{CE} for C and Eiffel.
Rather than directly replicating or wrapping commonly used external
libraries (such as \libname{stdio} and \libname{stdlib}), the helper classes
target C fundamental \emph{language features} and in particular types
and type constructors.  This approach works with any external library,
even non-standard ones, and is easier to maintain because involves 
only a limited number of classes.  We now give a concise description
of the most important helper classes; Section~\ref{sec:impl} shows
detailed examples of their usage.
\begin{itemize}
\item~$\!\!$\eif{CE_POINTER [G]} represents C pointers of any type
  through the generic parameter \eif{G}. It includes features to
  perform full-fledged pointer arithmetic and to get pointer
  representations which C can access but the Eiffel's runtime will not
  modify (in particular, the garbage collector will not modify pointed
  addresses nor relocate memory areas).

\item~$\!\!$\eif{CE_CLASS} defines the basic interface of Eiffel classes
  that correspond to \cc{union}s and \cc{struct}s. It includes features
  (members) that return instances of class \linebreak\eif{CE_POINTER} pointing to a
  memory representation of the structure that C can access.

\item~$\!\!$\eif{CE_ARRAY [G]} extends \eif{CE_POINTER} and provides
  consistent array access to both C and Eiffel (according to their
  respective conventions). It includes contracts that
  check for out-of-bound access.

\item~$\!\!$\eif{CE_ROUTINE} represents function pointers. It supports calls to Eiffel routines through \eif{agent}s---Eiffel's
  construct for function objects (\emph{closures} or
  \emph{delegates} in other languages)---and calls to (and callbacks from) external C functions through raw function pointers.

\item~$\!\!$\eif{CE_VA_LIST} supports variadic functions, using the Eiffel
  class \eif{TUPLE} (sequences of elements of heterogeneous type) to
  store a variable number of arguments.  It offers an Eiffel interface
  that extends the standard C's (declared in \headname{stdarg.h}), as
  well as output in a format accessible by external C code.
\end{itemize}



\section{Translating C to Eiffel}\label{sec:impl}

This section presents the major details of the translation \transl{}
from C to Eiffel implemented in \nameSys, and illustrates the general
rules with a number of small examples. The presentation breaks
down $\transl$ into several components that target different language
aspects (for example, $\tr{TD}$ maps C type declarations to Eiffel
classes); these components mirror the incremental transformations
$T_i$ of \nameSys (mentioned in Section~\ref{sec:principles}) but
occasionally overlook inessential details for presentation
clarity.

\titledpar{External functions in Eiffel.}
Eiffel code translated from C normally includes calls to external C pre-compiled functions, whose actual arguments correspond to objects in the Eiffel runtime.
This feature relies on the \eif{external} Eiffel language construct: Eiffel routines can be declared as \eif{external} and directly execute C code embedded as Eiffel strings\footnote{For readability, we will omit quotes in \eif{external} strings.} or call functions declared in header files.
For example, the following Eiffel routine (method) \eif{sin_twice} returns twice the sine of its argument by calling the C library function \cc{sin} (declared in \headname{math.h}):

\begin{codesnip}
\begin{lstlisting}[language=OOSC2Eiffel]
 sin_twice (arg: REAL_32): REAL_32
   external #\small\verb|C inline use <math.h>|# alias #\small\verb|return 2*sin($arg);|#  end
\end{lstlisting}
\end{codesnip}
Calls using \eif{external} can exchange arguments between the Eiffel and the C runtimes only for a limited set of primitive type: numeric types (that have the same underlying machine representation in Eiffel and C) and instances of the Eiffel system class \eif{POINTER} that corresponds to raw untyped C pointers (not germane to Eiffel's pointer representation, unlike \eif{CE_POINTER}).
In the \eif{sin_twice} example, argument \eif{arg} of numeric type \eif{REAL_32} is passed to the C runtime as \verb|$arg|.
Every helper class (described in Section~\ref{sec:principles}) includes an attribute \eif{c_pointer} of type \eif{POINTER} that offers access to a C-conforming representation usable in \eif{external} calls.

The mechanism of external calls is used not only in the translations produced by \nameSys but also in the Eiffel standard libraries, wherever interaction with the operating system is required, such as for input/output, file operations, and memory management.
Supporting calls to native code is also the only strict requirement to be able to implement, in languages other than Eiffel, this paper's approach to automatic translation from C.
Java, for example, offers similar functionalities through the Java Native Interface (JNI).

\subsection{Types and Type Constructors} \label{sec:types-type-constr}

C declarations \cc{T v$\,$} of a variable \cc{v} of type \cc{T} become Eiffel declarations \eif{v$\,$: $\tr{TY}$(T)}, where $\tr{TY}$ is the mapping from C types to Eiffel classes described in this section.

\titledpar{Numeric types.}  C numeric types correspond to Eiffel
classes \eif{INTEGER} (signed integers), \eif{NATURAL} (unsigned
integers), \eif{REAL} (floating point numbers) with the appropriate
bit-size as follows\footnote{We implemented class \eif{REAL_96} specifically to support \cc{long double} on Linux machines.}:

\begin{codesnip}
\begin{center}
\begin{tabular}{cc}
\textsc{C type} \cc{T}  &   \textsc{Eiffel class} \eif{$\tr{TY}(T)$} \\
\hline
\cc{char}  &  \eif{INTEGER_8} \\
\cc{short int}    &  \eif{INTEGER_16} \\
\cc{int, long int}    &  \eif{INTEGER_32} \\
\cc{long long int}  &  \eif{INTEGER_64} \\
\cc{float}   &  \eif{REAL_32} \\
\cc{double}  &  \eif{REAL_64} \\
\cc{long double} & \eif{REAL_96}
\end{tabular}
\end{center}
\end{codesnip}
Unsigned variants follow the same size conventions as signed integers
but for class \eif{NATURAL}; for example \cc{$\tr{TY}$(unsigned short int)} is \eif{NATURAL_16}.

\titledpar{Pointers.}
Pointer types are translated using class \eif{CE_POINTER [G]} with the generic parameter \eif{G} instantiated with the pointed type:

\begin{codesnip}
\begin{center}
\begin{tabular}{cc}
$\tr{TY}$\cc{(T *)}  & =  \eif{CE_POINTER [$\tr{TY}$(T)]}
\end{tabular}
\end{center}
\end{codesnip}
with the convention that $\tr{TY}$\cc{(void)} maps to Eiffel class
\eif{ANY}, ancestor to every other class (\eif{Object} in Java).  The
definition works recursively for multiple indirections; for example,
\eif{CE_POINTER[CE_POINTER[REAL_32]]} stands for $\tr{TY}$\cc{(float **)}.

\titledpar{Function pointers.}
Function pointers are translated to Eiffel using class \eif{CE_ROUTINE}:

\begin{codesnip}
\begin{center}
\begin{tabular}{c}
$\tr{TY}$\cc{($T_0\;$  (*)  $\;$($T_1$, ..., $T_n$))}  $\ $=$\ $ \eif{CE_ROUTINE}
\end{tabular}
\end{center}
\end{codesnip}
\eif{CE_ROUTINE} inherits from \eif{CE_POINTER [ANY]}, and hence it behaves as a generic pointer, but customizes its behavior using references to \eif{agent}s that wrap the functions pointed to;
Section~\ref{sec:vari-init-usage} describes this mechanism.

\titledpar{Arrays.}
Array types are translated to Eiffel using class \eif{CE_ARRAY[G]} with the generic parameter \eif{G} instantiated with the array base type: \cc{$\tr{TY}$(T [n])} = \eif{CE_ARRAY[$\tr{TY}$(T)]}.
The size parameter $n$, if present, does not affect the declaration, but initializations of array variables use it (see Section~\ref{sec:vari-init-usage}).
Multi-dimensional arrays are defined recursively as arrays of arrays: \cc{$\tr{TY}$(T [$n_1$][$n_2$]...[$n_m$])} is then \eif{CE_ARRAY[$\tr{TY}$(T[$n_2$]...[$n_m$])]}.

\titledpar{Enumerations.}
For every type \cc{enum E} defined or used, the translation introduces an Eiffel class \eif{E} defined by the translation $\tr{TD}$ (for type definition):

\begin{codesnip}
  $\quad\tr{TD}$\cc{(enum E $\;\{v_1 = k_1, \ldots, v_m = k_m\}$)} = \\
  \eif{$\qquad\qquad$class E feature}
\eif{$\;v_1$: INTEGER_32 = $k_1$; $\;\ldots\;$ ; $\;v_m$: INTEGER_32 = $k_m\,$ end}
\end{codesnip}

\noindent
Class \eif{E} has as many attributes as the \cc{enum} type has values, and each attribute is an integer that receives the corresponding value in the enumeration. Every C variable of type \cc{E} also becomes an integer variable in Eiffel (that is, $\tr{TY}$\cc{(enum E$\,$)} = \eif{INTEGER_32}), and the class \eif{E} is only used to assign constant values according to the \cc{enum} naming scheme.

\titledpar{Structs and unions.}
For every compound type \cc{struct S} defined or used, the translation introduces an Eiffel class \eif{S}: \\
\begin{codesnip}
    \eif{$\qquad\quad$  class S inherit CE_CLASS feature}
$\tr{F}(T_1\;v_1)  \ldots  \tr{F}(T_m\;v_m)$ \eif{end}
\end{codesnip}

\noindent
for $\tr{TD}$\cc{(struct S $\;\{T_1\;v_1; \ldots; T_m\;v_m\}$)}.
Correspondingly, $\tr{TY}$(\cc{S})$\,=\,$\eif{S }; that is, variables of type \cc{S} become references of class \eif{S} in Eiffel.
The translation $\tr{F}$(\cc{T v}) of each field \cc{v} of the \cc{struct S} introduces an attribute of the appropriate type in class \eif{S}, and a setter routine \eif{set_v} which also updates the underlying C representation of \cc{v}:

\begin{codesnip}
\begin{lstlisting}[language=OOSC2Eiffel]
  v: $\tr{TY}$(T) assign set_v     $\ $  -- declares `set_v' as the setter of v
  set_v (a_v: $\tr{TY}$(T)) do v := a_v ; update_memory_field ("v") end
\end{lstlisting}
\end{codesnip}

\noindent
Class \eif{CE_CLASS}, from which \eif{S} inherits, implements \eif{update_memory_field} using reflection, so that the underlying C representation is created and updated dynamically only when needed during execution (for example, to pass a \cc{struct} instance to a native C library), thus avoiding any data duplication overhead whenever possible.

\begin{example}
\label{ex:car-struct}
Consider a C \cc{struct car} that contains an integer field \cc{plate_num} and a string field \cc{brand}: 
\begin{codesnip}
\begin{lstlisting}[language=CustomC]
   typedef struct { unsigned int plate_num; char* brand; } car;
\end{lstlisting}
\end{codesnip}
The translation $\tr{TD}$ introduces a class \eif{CAR} as follows:
\begin{codesnip}
\begin{lstlisting}[language=OOSC2Eiffel]
   class CAR inherit CE_CLASS feature

      plate_num: NATURAL_32 assign set_plate_num

      brand: CE_POINTER [INTEGER_8] assign set_brand

      set_plate_num (a_plate_num: NATURAL_32)
         do plate_num := a_plate_num; update_memory_field ("plate_num") end

      set_brand (a_brand: CE_POINTER [INTEGER_8])
         do brand := a_brand; update_memory_field ("brand") end
   end
\end{lstlisting}
\end{codesnip}
\end{example}

The translation of \cc{union} types follows the same lines as that of \cc{struct}s, with the only difference that classes translating \cc{union}s generate the underlying C representation in any case upon initialization, even if the \cc{union} is not passed to the C runtime;
calls to \eif{update_memory_field} update all attributes of the class to reflect the correct memory value.
We found this to be a reasonable compromise between performance and complexity of memory management of \cc{union} types where, unlike \cc{struct}s, fields share the same memory space.

\subsection{Variable Initialization and Usage} \label{sec:vari-init-usage}
\noindent
\titledpar{Initialization.}
Eiffel variable declarations \eif{v$\,$: C} only allocate memory for a \emph{reference} to objects of class \eif{C}, and initialize it to \eif{Void} (\lstinline[language=Java]|null| in Java).
The only exceptions are, once again, numeric types: a declaration such as \eif{n: INTEGER_64} reserves memory for a 64-bit integer and initializes it to zero.
Therefore, every C local variable declaration \cc{T v} of a variable \cc{v} of type \cc{T} may also produce an \emph{initialization}, consisting of calls to creation procedures of the corresponding helper classes, as specified by the declaration mapping $\tr{DE}$:

\begin{center}
\begin{codesnip}
$\tr{DE}$\cc{(T v;)}  = $\!\!\!\!$
$\begin{cases}
v:\tr{TY}(T)       &        \text{(NT)}\\
v:\tr{TY}(T) \text{\eif{; create v.make}}(\ll \!n_1, \ldots, n_m\!\gg)  & \text{(AT)} \\
v: \tr{TY}(T) \text{\eif{; create v.make}}  &   \text{(OT)} \\
\end{cases}$
\end{codesnip}
\end{center}

\noindent
where definition (NT) applies if \eif{T} is a numeric type; (AT) applies if \eif{T} is an array type \eif{S[$n_1$], $\ldots$, [$n_m$]}; and (OT) applies otherwise.
The creation procedure \eif{make} of \eif{CE_ARRAY} takes a sequence of integer values to allocate the right amount of memory for each array dimension; for example \cc{int a[2][3]} is initialized by \eif{create a.make($\ll$2, 3$\gg$)}.

\titledpar{Memory management.}  Helper classes are regular Eiffel
classes; therefore, the Eiffel garbage collector disposes instances
when they are no longer referenced (for example, when a local variable
gets out of scope).  Upon collection, the \eif{dispose} finalizer
routines of the helper classes ensure that the C memory
representations are also appropriately deallocated; for example, the
finalizer of \eif{CE_ARRAY} frees the array memory area by calling
\cc{free} on the attribute \eif{c_pointer}.

To replicate the usage of \cc{malloc} and \cc{free}, we offer wrapper routines
that emulate the syntax and functionalities
of their C homonym functions, but operate on \eif{CE_POINTER}: they get raw C pointers by external calls to C library functions, convert them to \eif{CE_POINTER}, and record the dynamic information about allocated memory size.
The latter is used to check that successive usages conform to the declared size (see Section~\ref{sec:bugs}).
Finally, the creation procedure
\eif{make_cast} of the helper classes can convert a generic pointer returned by \eif{malloc} to the proper pointed type, according to the following translation scheme:

\begin{center}
\begin{tabular}{ll}
\textsc{C code}  &   \textsc{Translated Eiffel code} \\
\hline
\cc{T* p;}     &    \eif{p: CE_POINTER[$\tr{TY}$(T)]}\\
\cc{p = (T *) malloc(sizeof(T));}  &  \eif{create p.make_cast (malloc ($\sigma$(T)))} \\
\cc{free(p);}  &  \eif{free(p)}
\end{tabular}
\end{center}

\noindent
where $\sigma$ is an encoding of the size information.

\titledpar{Variable usage.}
The translation of variable usage is straightforward:
variable reads in expressions are replicated verbatim, and
C assignments (\cc{=}) become Eiffel assignments (\eif{:=}); the latter is,
for \eif{CE_ARRAY}, \eif{CE_POINTER}, and classes translating C \cc{struct}s and \cc{union}s,
syntactic sugar for calls to setter routines that achieve the desired effect.
The only exceptions occur when implicit type conversions in C must become explicit in Eiffel, which may spoil the readability of the translated code but is necessary with strong typing.
For example, the C assignment \cc{cr = 's'}---assigning character constant \cc{'s'} to variable \cc{cr} of type \cc{char}---becomes the Eiffel assignment \eif{cr := ('s').code.to_integer_8} that encodes \cc{'s'} using the proper representation.

\titledpar{Variable address.}
Whenever the address \cc{&v} of a C variable \cc{v} of type \cc{T} is taken, \cc{v} is translated as an array of unit size and type \cc{T}:
\begin{codesnip}
$\tr{DE}$\cc{(T v)} = $\tr{DE}$\cc{(T v[1])},
\end{codesnip}
and every usage of \cc{v} is adapted accordingly: \cc{&v} becomes just \cc{v}, and occurrences of \cc{v} in expressions become \cc{*v}.
This little hack makes it possible to have Eiffel assignments translate C assignment uniformly; otherwise, usages of \cc{v} should have different translations according to whether the information about \cc{v}'s memory location is copied around (with \cc{&}) or not.

\titledpar{Dereferencing and pointer arithmetic.}
The helper class \eif{CE_POINTER} features a query \eif{item} which translates dereferencing (\cc{*}) of C pointers.
Pointer arithmetic is translated verbatim, because class \eif{CE_POINTER} overloads the arithmetic operators to be aliases of proper underlying pointer manipulations, so that an expression such as \eif{p + 3} in Eiffel, for references \eif{p} of type \eif{CE_POINTER}, hides the explicit expression \eif{c_pointer + 3 * element_size}.

\begin{example}
\label{ex:car-pointers}

Consider an integer variable \cc{num}, a pointer variable \cc{p}, and a double pointer variable \cc{carsp}  with target type \cc{struct car} (defined in Example~\ref{ex:car-struct}).
The following C code fragment declares \cc{num}, \cc{p}, and \cc{carsp}, allocates space for an array
of \cc{num} consecutive cars pointed to by \cc{carsp}, and makes \cc{p} point to the array's third element.

\begin{codesnip}
\begin{lstlisting}[language=CustomC]
   int num;
   car * p;
   car ** carsp;
   *carsp = malloc(num * sizeof(car));
   p = *carsp + 2;
\end{lstlisting}
\end{codesnip}

\noindent
\nameSys translates the C fragment as follows, where the implicit C conversion from signed to unsigned \cc{int} become explicit in Eiffel (calls to routine \eif{to_natural_32}).

\begin{codesnip}
\begin{lstlisting}[language=OOSC2Eiffel]
   num: INTEGER_32
   p: CE_POINTER [CAR]
   carsp: CE_POINTER [CE_POINTER [CAR]]
   carsp.item := create {CE_POINTER [CAR]}.make_cast
         (malloc (num.to_natural_32 * (create {CAR}.make).structure_size.to_natural_32))
   p := carsp.item + 2
\end{lstlisting}
\end{codesnip}
\end{example}
%

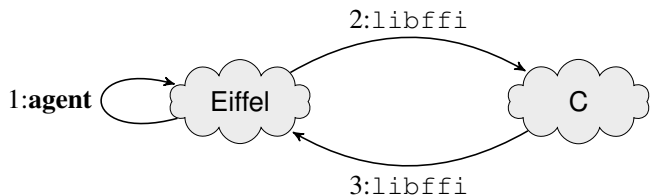
\begin{figure}
\tikzsetnextfilename{fpointers}
\centering
\begin{tikzpicture}[->,>=stealth',shorten >=1pt,auto,node distance=4.5cm,semithick]
\tikzstyle{every state}=[cloud,draw,fill=gray!15,aspect=2.5,very thin]
\node[state] (eif) {\textsf{Eiffel}};
\node[state] (cc) [right of=eif] {\textsf{C}};
\path (eif) edge [loop left] node {1:\eif{agent}} (eif);
\path (eif) edge [bend left] node {2:\libname{libffi}} (cc);
\path (cc) edge [bend left] node {3:\libname{libffi}} (eif);
\end{tikzpicture}
\caption{Calls through function pointers.}\label{fig:fps}
\end{figure}

\titledpar{Using function pointers.}
Class \eif{CE_ROUTINE}, which translates C function pointers, is usable both in the Eiffel and in the C environment (see Figure~\ref{fig:fps}).
On the Eiffel side, its instances wrap Eiffel routines using \emph{agents}---Eiffel's mechanism for function objects.
A private attribute \eif{routine} references objects of type \eif{ROUTINE [ANY, TUPLE]}, an Eiffel system class that corresponds to agents wrapping routines, with any number of arguments and argument types stored in a tuple.
Thus, Eiffel code can use the \eif{agent} mechanism to create instances of class \eif{ROUTINE}.
For example, if \eif{foo} denotes a routine of the current class and \eif{fp} has type \eif{CE_ROUTINE}, \eif{create fp.make_agent (agent foo)} makes \eif{fp}'s attribute \eif{routine} point to \eif{foo}.
On the C side, when function pointers are directly created from C pointers (e.g., references to external C functions), \eif{CE_ROUTINE} behaves as a wrapper of raw C function pointers, and dynamically creates invocations to the pointed functions using the library \libname{libffi}~\cite{libffi}.

The Eiffel interface to \eif{CE_ROUTINE} will then translate calls to wrapped functions into either \eif{agent} invocations or external calls with \libname{libffi} according to how the class has been instantiated.
Assume, for example, that \eif{fp} is an object of class \eif{CE_ROUTINE} that wraps a procedure with one integer argument.
If \eif{fp} has been created with an Eiffel \eif{agent foo} as above, calling \eif{fp.call ([42])} wraps the call \eif{foo (42)}  (edge 1 in Figure~\ref{fig:fps}); if, instead, \eif{fp} only maintains a raw C function pointer, the same instruction \eif{fp.call ([42])} creates a native C call using \libname{libffi} (edge 2 in Figure~\ref{fig:fps}).

The services of class \eif{CE_ROUTINE} behave as an adapter between the procedural and object-oriented representation of routines: the signatures of C functions must change when they are translated to Eiffel routines, because routines in object-oriented languages include references to a target object as implicit first
argument.  Calls from external C code to Eiffel routines are therefore
intercepted at runtime with \libname{libffi} callbacks (edge 3 in
Figure~\ref{fig:fps}) and dynamically converted to suitable
\eif{agent} invocations.

\subsection{Conditionals, Loops, and Functions}
\noindent
The translation $\tr{CF}$ takes care of constructs to structure the control flow of computations: conditionals, loops, and function definitions and calls.

\titledpar{Conditionals and loops.}
Sequential composition, conditional instructions, and loop instructions
are similar in C and Eiffel; therefore, their translation is straightforward.\footnote{Eiffel loops have the general form: \eif{from init until exit loop B end}. \eif{init} is executed once before the actual loop; the loop body \eif{B} is executed until the Boolean condition \eif{exit} becomes true; since the exit condition is tested before each iteration, the body may not be executed. Body \eif{B} and \eif{init} may be empty.}

\begin{center}
\begin{tabular}{ll}
$\tr{CF}( I_1 \,;\, I_2)$  & = \quad $\tr{}(I_1)\,;\,\tr{}(I_2)$ \\
$\tr{CF}$\cc{(if (c) $\,$\{TB\} else \{EB\})}  & =
\quad \eif{if $\;\tr{}$(c)} \\ &  \!\quad $\quad$\eif{then $\,\tr{}$(TB) $\;$else $\,\tr{}$(EB) $\,$end} \\
$\tr{CF}$\cc{(while (c) $\,$\{LB\})}  & =
\quad \eif{from until not $\;\tr{}$(c)} \\
& \!\quad $\quad$\eif{loop $\,\tr{}$(LB) $\,$end}
\end{tabular}
\end{center}

Note that Eiffel has only one type of loop, which translates the unique type of C loops produced by CIL.

\titledpar{Functions.}
Function declarations and function calls in C translate to routine declarations and routine calls in Eiffel.
Given a function \cc{foo} with arguments $a_1, \ldots, a_n$ of type $T_1, \ldots, T_n$ and return type
$T_0$, the translation:
\begin{center}
\begin{codesnip}
$\tr{CF}\big(\;$\cc{$T_0\:$ foo ($T_1\ a_1,\ldots,T_n\ a_n$)\{ B \}}$\;\big)$
\end{codesnip}
\end{center}
is defined as follows:
\noindent
\begin{center}
\begin{codesnip}
$\begin{cases}
\text{\eif{foo}} (a_1\!\!:\! \tr{TY}(T_1);\ldots; a_n\!\!:\!\tr{TY}(T_n)) \;\text{\eif{do}} \,\tr{}(B)\, \text{\eif{end}}   & \text{if }T_0 \text{ is } \cc{void} \\
\text{\eif{foo}} (a_1\!\!:\! \tr{TY}(T_1);\ldots; a_n\!\!:\!\tr{TY}(T_n)): \tr{TY}(T_0) \;\text{\eif{do}} \,\tr{}(B)\, \text{\eif{end}}   & \text{otherwise}      \\
\end{cases}$
\end{codesnip}
\end{center}

Correspondingly, function calls in C become routine calls in Eiffel with the actual arguments also recursively translated:
\begin{center}
\begin{codesnip}
$\tr{CF}$(\cc{foo ($e_1, \ldots, e_n$)})  \quad =\quad \eif{foo ($\tr{}(e_1), \ldots, \tr{}(e_n)$)}
\end{codesnip}
\end{center}

\titledpar{Variadic functions.}
The C language supports \textit{variadic functions}, also known as \textit{varargs functions}, which work on a variable number of arguments.
The Eiffel translation of variadic function declarations uses the class \eif{TUPLE} to wrap lists of arguments:
the translation of a variadic function \cc{var_foo} with $n \geq 0$ required arguments $a_1, \ldots, a_n$ of type $T_1, \ldots, T_n$, optional additional arguments, and return type $T_0$:
\begin{center}
\begin{codesnip}
$\tr{CF}\big(\;$\cc{$T_0\:$ var_foo ($T_1\ a_1,\ldots,T_n\ a_n, \ldots$)\{ B \}}$\;\big)$
\end{codesnip}
\end{center}
is defined as:
\begin{center}
\begin{codesnip}
\eif{var_foo (args: TUPLE[$a_1\!: \tr{TY}(T_1); \;\ldots;\; a_n\!: \!\tr{TY}(T_n)$]): $\ \tr{TY}(T_0)$}
\end{codesnip}
\end{center}
assuming $T_0$ is not \cc{void}; otherwise, the return type is omitted, as in the translation of standard functions.

Calls to variadic functions with $n$ required arguments may provide $m \geq 0$ additional actual arguments $e_{n+1}, \ldots, e_{n+m}$ after the $n$ required ones $e_1, \ldots, e_n$.
The translation combines all arguments, required and additional, into an $(n+m)$-\eif{TUPLE} denoted by square brackets:
\begin{center}
\begin{codesnip}
$\tr{CF}$(\cc{var_foo ($e_1, \ldots, e_n, e_{n+1}, \ldots, e_{n+m}$)})  =
\eif{var_foo ([$\tr{}(e_1), \ldots, \tr{}(e_n), \tr{}(e_{n+1}), \ldots, \tr{}(e_{n+m})$])}
\end{codesnip}
\end{center}
The correctness of this translation relies on Eiffel's type conformance rule for instances of the \eif{TUPLE} class: every $(n+m)$-\eif{TUPLE} with types $[T_1, \ldots, T_n, T_{n+1},$ $\ldots, T_{n+m}]$, for $n, m \geq 0$, conforms to any shorter $n$-\eif{TUPLE} with types $[T_1, \ldots, T_n]$.
Thus, translated calls are type-safe because the assignment of actual to formal arguments is covariant~\cite{GJ-PL}.

The bodies of variadic functions can refer to the required arguments by name using the usual syntax.
The C type \cc{va_list} and functions \cc{va_start} and \cc{va_arg} in library \libname{stdarg} provide a means to access the optional arguments: \cc{va_start(argp, last)} initializes a variable \cc{argp} of type \cc{va_list} to point to the first optional argument after \cc{last} (the name of the last declared argument).
Every successive invocation of \cc{va_arg(argp, T)} returns the argument of type \cc{T} pointed to by \cc{argp}, and then moves \cc{argp} right after it.
The Eiffel translation relies on the helper class \eif{CE_VA_LIST} which replicates \libname{stdarg}'s functionality:
\begin{center}
\begin{codesnip}
\begin{tabular}{ll}
$\tr{DE}$( \cc{va_list argp} )  & = \quad \eif{argp: CE_VA_LIST}  \\
$\tr{}$( \cc{va_start(argp, last)} )  & = \quad \eif{create argp.make (args, index)} \\
$\tr{}$( \cc{va_arg(argp, T)} ) & = \quad \eif{argp.$\tr{TY}(T)$_item}
\end{tabular}
\end{codesnip}
\end{center}
where \cc{args} is the name of the argument \eif{TUPLE} in the enclosing variadic function and \cc{index} is the index of the first optional argument.

While \cc{CE_VA_LIST} provides a uniform interface to both Eiffel and C, the actual arguments of a variadic function may also be accessed in the Eiffel translation using the standard syntax for \eif{TUPLE}s: in a variadic function with $n$ required arguments, the expression \eif{args.$a_i$} refers to the $i$th named argument $a_i$, for $1 \leq i \leq n$; and \eif{args.$T_{j}$_item(j)} refers to the $j$th argument $a_j$, for any $1 \leq j \leq n$ (required argument) or $j > n$ (optional argument), where $T_{j}$ is $a_j$'s type.

\begin{example}
\label{ex:varargs}
Continuing Examples~\ref{ex:car-struct} and~\ref{ex:car-pointers}, consider a variadic function \cc{init_cars} that takes a pointer \cc{carsp} to an array of cars and \cc{num} pairs of unsigned integers and strings (passed as optional arguments), and initializes the array with \cc{num} cars whose plate numbers and brands respectively correspond to the integer and string in each pair.

\begin{codesnip}
\begin{lstlisting}[language=CustomC]
   void init_cars(car **carsp, int num, $\ldots$) {
      va_list argp;
      car *ccar;
      int n;
      *carsp = malloc(num * sizeof(car));
      ccar = *carsp;
      va_start(argp, num);
      for(n = num ; n > 0; n--) {
         ccar$\,\rightarrow$plate_num = va_arg(argp, unsigned int);
         ccar$\,\rightarrow$brand = va_arg(argp, char*);
         ccar++;
      }
   }
\end{lstlisting}
\end{codesnip}

The translation into Eiffel uses the class \eif{CE_VA_LIST} and accesses the optional unsigned integer arguments with \eif{argp.natural_32_item}, and the optional strings with \eif{argp.pointer_item}.
The latter returns a generic pointer, which is cast to \cc{char *} with an explicit cast (line~\ref{l:argp-pointer-cast}).
\pagebreak

\begin{codesnip}
\begin{lstlisting}[language=OOSC2Eiffel,numbers=left,numberstyle=\tiny, numbersep=0.5pt, stepnumber=15]
init_cars (args: TUPLE [carsp: CE_POINTER [CE_POINTER [CAR]]; num: INTEGER_32])
 local
    argp: CE_VA_LIST
    ccar: CE_POINTER [CAR]
    n: INTEGER_32
 do
    args.carsp.item := create {CE_POINTER [CAR]}.make_cast (
          malloc (  num.to_natural_32 * (create {CAR}.make).structure_size.to_natural_32 )
         )
    ccar := args.carsp.item
    create argp.make (args, 3)
    n := num
    from until not n > 0 loop
       ccar.item.plate_num := argp.natural_32_item
       ccar.item.brand := (create {CE_POINTER [INTEGER_8]}.make_cast (argp.pointer_item)) #\label{l:argp-pointer-cast}#
       ccar := ccar + 1
       n := n - 1
    end
 end
\end{lstlisting}
\end{codesnip}
\end{example}

\subsection{Unstructured Control Flow}
\noindent
In addition to constructs for structured programming, C offers control-flow breaking instructions such as jumps.
This section discusses their translation to Eiffel.

\titledpar{Jumps.}
Eiffel enforces structured programming, and hence it lacks control-flow breaking instructions such as \cc{goto}, \cc{break}, \cc{continue}, and \cc{return}.
The translation $\tr{CF}$ eliminates such instructions along the lines of the \emph{global} version---using Harel's terminology~\cite{Harel-folk}---of the structured programming theorem.
The body of a function using \cc{goto} determines a list of instructions $s_0, s_1, \ldots, s_n$, where each $s_i$ is a maximal sequential block of instructions, with no labels after the first instruction or jumps before the last one in the block.
$\tr{CF}$ translates the body into a single loop over an auxiliary integer variable \eif{pc} that emulates a \emph{program counter}:\footnote{Eiffel's \eif{inspect/when} instructions corresponds to a restricted form of \cc{switch/case} without fall through.}

\begin{codesnip}
$$
\tr{CF}(\langle s_0, s_1, \ldots, s_n\rangle) =
\left\{ \begin{array}{l}
\text{\eif{from pc := 0 until pc = -1 loop}} \\
\ \text{\eif{inspect pc}} \\
\ \ \text{\eif{when 0 then}}\,\tr{}(s_0)\,;\text{\eif{upd(pc)}} \\
\ \ \text{\eif{when 1 then}}\,\tr{}(s_1)\,;\text{\eif{upd(pc)}} \\
\ \ \vdots \\
\ \ \text{\eif{when n then}}\,\tr{}(s_n)\,;\text{\eif{upd(pc)}} \\
\ \text{\eif{end}} \\
\text{\eif{end}}
\end{array} \right.
$$
\end{codesnip}

Variable \eif{pc} is initially zero; every iteration of the loop body executes $s_{\text{\eif{pc}}}$ for the current value of \eif{pc}, and then updates \eif{pc} (with \eif{upd(pc)}) to determine the next instruction to be executed: blocks ending with jumps modify \eif{pc} directly, other blocks increment it by one, and exit blocks set it to $-1$, which makes the overall loop terminate (whenever the original function terminates).

This translation supports all control-flow breaking instructions, and in particular \cc{continue}, \cc{break}, and \cc{return}, which are special cases of \cc{goto}.
$\tr{CF}$, however, improves the readability in these special simpler cases by directly using auxiliary Boolean flag variables with the same names as the instruction they replace.
The flags are tested as part of the translated exit conditions for the loops where the control-flow breaking instructions appear.
Using this alternative translation scheme where the generality of \cc{goto}s is not required makes for translations with little changes to the code structure, which are usually more readable.
The loop \cc{while (n > $\,$0) \{ if (n == 3) $\:$break; n--; \}}, for example, becomes:

\begin{codesnip}
\begin{lstlisting}[language=OOSC2Eiffel]
    from until break or not n > 0 loop
      if n = 3 then break := True end
      if not break then n := n - 1 end
    end
\end{lstlisting}
\end{codesnip}

\titledpar{Long jumps.}  The C library \libname{setjmp} provides
functions \cc{setjmp} and \cc{longjmp} to save an arbitrary return
point and jump back to it across function call boundaries.  The
wrapping mechanism used for external functions (see
Section~\ref{sec:externals}) does not work to replicate long jumps,
because the return values saved by \cc{setjmp} wrapped as an external
function are no longer valid after execution leaves the
wrapper. Therefore, \nameSys translates \cc{setjmp} and \cc{longjmp}
by means of the helper class \eif{CE_EXCEPTION}.  As the name
suggests, \eif{CE_EXCEPTION} uses Eiffel's exception propagation
mechanism to go back in the call stack to the allocation frame of the
function that called \cc{setjmp}. There, translated \cc{goto}
instructions jump to the specific point saved with \cc{setjmp} within
the function body.

\begin{example}
\label{ex:longjmp}
Consider a C function \cc{target} that executes \cc{setjmp(buf)} to save a location using a global variable \cc{buf} of type \cc{jmp_buf}.
In the translated Eiffel code, \cc{buf} becomes an attribute of type \cc{CE_EXCEPTION}, and the call to \cc{setjmp} in \cc{target} becomes a creation of an instance attached to \eif{buf}: \eif{buf := create \{CE_EXCEPTION\}.make ($\ldots$)}; \eif{make}'s actual arguments (omitted for simplicity) contain references to the stack frame of the current instance of \eif{target}, and to the specific instruction where to jump.

Later during the execution, another function \cc{source} called from \cc{target} performs a \cc{longjmp(buf, 3)}, which diverts execution to the location previously marked by \cc{setjmp} and returns value \cc{3}.
In Eiffel, the \cc{longjmp} becomes a raising of the exception object \eif{buf} enclosing the return value: \eif{buf.raise (3)}.

The flow of what happens next depends on the semantics of Eiffel exceptions, which is significantly different than that of other object-oriented languages such as Java and C\#.
The raised exception propagates to the recipient \eif{target}, where it is handled by a \eif{rescue} clause.
Such exception handling blocks are routine-specific in Eiffel, rather than being associated to arbitrary scopes such as in Java's \lstinline[language=Java]|try/catch| blocks.
The translation of \cc{setjmp} also took care of setting up \eif{target}'s \eif{rescue} clause: if the raised exception stores a reference to the current instance of routine \eif{target}, the code in the \eif{rescue} clause modifies a variable \eif{pc} local to \eif{target} to point to the location of the \cc{setjmp} and, using Eiffel's \eif{retry} instruction, executes \eif{target}'s body again with this new value.
Based on the value, the new execution of \eif{target}'s body jumps to the correct location following a mechanism similar to the previously discussed translation of \cc{goto} instructions.
Finally, if the handled exception references a routine instance other than the current execution of \eif{target}, the \eif{rescue} clause propagates the exception through the call stack, so that it can reach the correct stack frame.
\end{example}

\subsection{Externals and Encapsulation} \label{sec:externals}
\noindent
The translation $\tr{}$ uses classes to wrap header and source files, which give a simple modular structure to C programs.

\titledpar{Externals.}
For every included system header \headname{header.h}, $\transl$ defines a class \eif{HEADER} with wrappers for all external functions and variables declared in \headname{header.h}.
The wrappers are routines using the Eiffel \eif{external} mechanism and performing the necessary conversions between the Eiffel and the C runtimes.
In particular, external functions using only numeric types, which are interoperable between C and Eiffel, directly map to wrapper routines; for example, \cc{exit} in \headname{stdlib.h} becomes:

\begin{codesnip}
\begin{lstlisting}[language=OOSC2Eiffel]
    exit (status: INTEGER_32)  external #\small\verb|C inline use <stdlib.h>|#
                            $\:$alias #\small\verb|exit($status);|# end
\end{lstlisting}
\end{codesnip}
When external functions involve types using helper classes in Eiffel, a routine passes the underlying C representation to the external calls; for example, \cc{fclose} in \headname{stdio.h} generates:

\begin{codesnip}
\begin{lstlisting}[language=OOSC2Eiffel]
   fclose (stream: CE_POINTER [ANY]): INTEGER_32
        do Result := c_fclose (stream.c_pointer) end

   c_fclose (stream: POINTER): INTEGER_32
        external #\small\verb|C inline use <stdio.h>|#
        alias #\small\verb|return fclose($stream);|# end
\end{lstlisting}
\end{codesnip}
In some complex cases---typically, with variadic external functions---the wrapper can only assemble the actual call on the fly at runtime; this is done using \eif{CE_ROUTINE}.
For example, \cc{printf} is wrapped as:

\begin{codesnip}
\begin{lstlisting}[language=OOSC2Eiffel]
  printf (args: TUPLE[format: CE_POINTER[INTEGER_8]]): INTEGER_32
  do
   Result := (create {CE_ROUTINE}.make_shared (c_printf)).integer_32_item (args)
  end

  c_printf: POINTER
     external #\small\verb|C inline use <stdio.h>|# alias #\small\verb|return &printf;|# end
\end{lstlisting}
\end{codesnip}

The translation can also inline \textbf{assembly code}, using the same mechanisms as external function calls.

\titledpar{Globals.}
For every source file \headname{source.c}, $\transl$ defines a class \eif{SOURCE} that includes translations of all function definitions (as routines) and global variables (as attributes) in \headname{source.c}.
Class \eif{SOURCE} also \emph{inherits} from the classes translating the other system header files that \headname{source.c} includes, to have access to their declarations.
For example, if \headname{foo.c} includes \headname{stdio.h}, \eif{FOO} is declared as \eif{class FOO inherit STDIO}.

\subsection{Formatted Output Optimization} \label{sec:form-outp-optim}

The library function \cc{printf} is the standard C output function, which
displays values according to format strings passed as argument.
In contrast, Eiffel provides the command \eif{Io.put_string} to put plain text strings on
standard output, as well as type-specific formatter classes, such as \eif{FORMAT_INTEGER},
to produce string representations of arbitrary types.

With the goal of making the translated code as
close as possible to standard Eiffel, \nameSys tries to replace calls to
\cc{printf} with equivalent calls to \eif{Io.put_string} and routines of the formatter classes.
Whenever \cc{printf}'s returned value is not
used and the format string is a constant literal, \nameSys parses the literal
format string and encodes it as equivalent calls to
\eif{Io.put_string} and formatters.
This process may find mismatches between some format
specifiers and the types of the corresponding value arguments, which \nameSys reports as warnings.
Another situation where \nameSys can replace calls to \cc{printf} with calls to \eif{put_string} is when only one
argument is passed to the former, which is then interpreted verbatim as a string.
Whenever a warning is issued or the usage of \cc{printf} cannot be rendered using \eif{put_string}, \nameSys falls back to calling a wrapper of the native \cc{printf} function (described in Section~\ref{sec:externals}).
The translation also implements similar optimizations for variants of \cc{printf} such as \cc{fprintf}.




\section{Evaluation and Discussion}\label{sec:eval}

This section evaluates the translation $\transl$ and its implementation in \nameSys.
The bulk of the evaluation assesses correctness (Section~\ref{sec:behavior}) and performance (Section~\ref{sec:performance}) based on experiments with 14 open-source programs.
We also qualitatively discuss other aspects that influence maintainability (Section~\ref{sec:readability}), the advantages in terms of safety deriving from switching to the Eiffel runtime (Section~\ref{sec:bugs}), as well as the current limitations of the \nameSys approach (Section~\ref{sec:limitations}).

\subsection{Correct Behavior} \label{sec:behavior}
To experimentally assess the translations produced by \nameSys, we applied it to 14 open-source C programs, including 7 applications, 6 libraries, and one testsuite; most of them are widely-used in Linux and other ``*nix'' distributions.
\progname{hello world} is the only toy application, which is however useful to have a baseline of translating from C to Eiffel with \nameSys.
The other applications are:
\progname{micro httpd} 12dec2005, a minimal HTTP server;
\progname{xeyes} 1.0.1, a widget for the X Windows System which shows two googly eyes following the cursor movements;
\progname{less} 382-1, a text terminal pager;
\progname{wget} 1.12, a command-line utility to retrieve content from the web;
\progname{links} 1.00, a simple web browser;
\progname{vim} 7.3, a powerful text editor.
The libraries are:
\progname{libSDL\_mixer} 1.2, an audio playback library;
\progname{libmongoDB} 0.6, a library to access MongoDB databases;
\progname{libpcre} 8.31, a library for regular expressions;
\progname{libcurl} 7.21.2, a URL-based transfer library supporting protocols such as FTP and HTTP;
\progname{libgmp} 5.0.1, for arbitrary-precision arithmetic;
\progname{libgsl} 1.14, a powerful numerical library.
The \progname{gcc} ``torture tests'' are short but semantically complex pieces of C code, used as regression tests for the GCC compiler.

Table~\ref{tab:eval} shows the results of translating the 14 programs into Eiffel using \nameSys, running on a GNU/Linux box (kernel 2.6.37) with a 2.66~GHz Intel dual-core CPU and 8~GB of RAM, GCC 4.5.1, CIL 1.3.7, EiffelStudio 7.1.8.
For each application, library, and testsuite Table~\ref{tab:eval} reports: (1)~the size (in lines of code) of the
CIL version of the C code and of the translated Eiffel code; (2)~the number of Eiffel classes created; (3)~the time (in seconds) spent by \nameSys to perform the source-to-source translation (not including compilation from Eiffel source to binary); (4)~the size of the binaries (in MBytes) generated by EiffelStudio.\footnote{We do not give a binary size for libraries, because EiffelStudio cannot compile them without a client.}


\begin{table}
\caption{Translation of 14 open-source programs.}\label{tab:eval}
\centering
\begin{tabular}{l r r r r r D{.}{.}{1}}
\toprule
            & \multicolumn{2}{c}{\textsc{Size (LOCS)}} & \textsc{\#Eiffel} & \textsc{Time} & \multicolumn{1}{c}{\textsc{Binary size}} \\
            &    \textsc{CIL}   &    \textsc{Eiffel}   & \textsc{classes}  &     \textsc{(s)}     &    \multicolumn{1}{c}{\textsc{(MB)}}     \\
\midrule
\progname{hello world}    &        8  &         15  &      1  &    1  &    1.3  \\
\progname{micro httpd}    &      565  &      1,934  &     16  &    1  &    1.5  \\
\progname{xeyes}          &    1,463  &     10,661  &     78  &    1  &    1.8  \\
\progname{less}           &   16,955  &     22,545  &     75  &    5  &    2.6  \\
\progname{wget}           &   46,528  &     57,702  &    183  &   25  &    4.5  \\
\progname{links}          &   70,980  &    100,815  &    211  &   33  &   13.9  \\
\progname{vim}            &  276,635  &    395,094  &    663  &  144  &   24.2  \\
\hline
\progname{libSDL\_mixer}  &    7,812  &     11,553  &     47  &    3  &     --  \\
\progname{libmongoDB}     &    7,966  &     10,341  &     43  &    3  &     --  \\
\progname{libpcre}        &   18,220  &     24,885  &     38  &   14  &     --  \\
\progname{libcurl}        &   37,836  &     65,070  &    289  &   18  &     --  \\
\progname{libgmp}         &   61,442  &     79,971  &    370  &   21  &     --  \\
\progname{libgsl}         &  238,080  &    344,115  &    978  &   85  &     --  \\
\hline
\progname{gcc} (torture)  &  147,545  &    256,246  &  2,569  &   79  &  1,576  \\
\hline
\textsc{Total}            &  932,035  &  1,380,947  &  5,561  &  433  &  1,626  \\
\bottomrule
\end{tabular}
\end{table}

We ran extensive trials on the translated programs to verify that they behave as in their original C version, thus validating the correctness of the translation \transl{} and its implementation in \nameSys.
The trials comprised informal usage, systematic performance tests for some of the applications, and running the standard testsuites available for the three biggest libraries (\progname{libcurl}, \progname{libgmp}, and \progname{libgsl}).
The rest of this section describes the translated testsuites and their behavior with respect to correctness; Section~\ref{sec:performance} discusses the quantitative performance results.

Library \progname{libcurl} comes with a client application and a testsuite of 583 tests defined in XML and executed by a Perl script calling the client; \progname{libgmp} and \progname{libgsl} respectively include testsuites of 145 and 46 tests, consisting of client C code using the libraries.
All tests execute and pass on both the C and the translated Eiffel versions of the libraries, with the same logged output. For \progname{libcurl}, \nameSys translated the library and the client application. For \progname{libgmp} and \progname{libgsl}, it translated the test cases as well as the libraries.

The \progname{gcc} torture testsuite includes 1116 tests; the GCC version we used fails 4 of them; CIL (which depends on GCC) fails another 110 tests among the 1112 that GCC passes; finally, \nameSys (which depends on CIL) passes 989 (nearly 99\%) and fails 13 of the 1002 tests passed by CIL.
Given the challenging nature of the torture testsuite, this result is strong evidence that \nameSys handles the complete C language used in practice, and produces correct translations.

The 13 torture tests failing after translation to Eiffel target the following unsupported features.
One test reads an \cc{int} from a \cc{va_list} (variadic function list of arguments) which actually stores a \cc{struct} whose first field is a \cc{double}; the Eiffel type-system does not allow this, and inspection suggests that it is probably a copy-paste error rather than a feature.
Two tests exercise GCC-specific optimizations, which are immaterial after translation to Eiffel.
Six tests target exotic GCC built-in functions, such as \cc{builtin_frame_address}; one test performs explicit function alignment; and three rely on special bitfield operations.


\begin{table}
\caption{Performance comparison for 5 applications and 4 testsuites.}\label{tab:eval2}
\centering
\begin{tabular}{l |rrr| rrr| rrr}
\toprule
           & \multicolumn{3}{c|}{\textsc{Execution time (s)}} & \multicolumn{3}{c|}{\textsc{Max \% CPU}} & \multicolumn{3}{c}{\textsc{Max MB RAM}} \\
            & C & T & E & C & T & E & C & T & E    \\
\midrule
\progname{hello world}   &   0 &     0 &  0     &  0  & 30 & 30       & 1.3 & 5.5 & 5.3     \\
\progname{micro httpd}   &   5 &    37 & 46     & 99  & 99 & 99       & 2.3 & 7.8 & 5.6     \\
\progname{less}          &  36 &    36 & --     & --  & -- & --       & --  & --  & --      \\
\progname{wget}          &  16 &    16 & --     & 22  & 22 & --       & 4.4 & 69  & --      \\
\progname{vim}           &  85 &    85 & --     & --  & -- & --       & --  & --  & --      \\
\hline
\progname{libcurl}       & 199 &   212 & --     &  -- & -- & --       & --  & --  & --      \\
\progname{libgmp}        &  44 &   728 & --     &  -- & -- & --       & --  & --  & --      \\
\progname{libgsl}        &  25 &  1501 & --     &  -- & -- & --       & --  & --  & --      \\
\progname{gcc} (torture) &   0 &     5 & --     &  -- & -- & --       & --  & --  & --      \\
\bottomrule
\end{tabular}
\end{table}

\subsection{Performance} \label{sec:performance}
\noindent
We analyzed the performance of 5 applications, the GCC torture testsuite, and the standard testsuites of 3 libraries (described in Section~\ref{sec:behavior}) translated to Eiffel using \nameSys.
Table~\ref{tab:eval2} shows the result of the performance trials, running on the same system used for the experiments of Section~\ref{sec:behavior}.\footnote{We compiled all CIL-processed C programs with the GCC options \mbox{\tt -O2 -s}; and all Eiffel programs with disabled void checking, inlining size 100, and stripped binaries.}
For each program or testsuite, Table~\ref{tab:eval2} reports the execution time (in seconds), the maximum percentage of CPU, and the maximum amount of RAM (in MBytes) used while running.
The table compares the performance of the original C versions (columns ``C'') against the Eiffel translations with \nameSys (columns ``T''), and, for the simpler examples, against manually written Eiffel implementations (columns ``E'') that transliterate the original C implementations using the closest Eiffel constructs (for example, \cc{putchar} becomes \eif{Io.put_character}) with as little changes as possible to the code structure (we manually wrote these transliterations ourselves).
Maximum CPU and RAM usages are immaterial for the testsuites (torture and libraries), because their execution consists of a large number of separate calls.
The rest of this section discusses the performance results together with other qualitative performance assessments.

The performance of \progname{hello world} demonstrates the base overhead, in terms of CPU and memory usage, of the default Eiffel runtime (objects, automatic memory management, and contract checking---which can however be disabled for applications where sheer performance is more important than having additional checks).

The test with \progname{micro httpd} consisted in serving the local download of a 174 MB file (the Eclipse IDE); this test boils down to a very long sequence (approximately 200 million iterations) of inputting a character from file and outputting it to standard output.
The translated Eiffel version incurs a significant overhead with respect to the original C version, but is faster than the manually written Eiffel transliteration.
This might be due to feature lookups in Eiffel or to the less optimized implementation of Eiffel's character services.
As a side note, we did the same exercise of manually transliterating \progname{micro httpd} using Java's standard libraries; this Java translation ran the download example in 170 seconds, using up to 99\% of CPU and 150 MB of RAM.

The test with \progname{wget} downloaded the same 174 MB Eclipse package over the SWITCH Swiss network backbone.
The bottleneck is the network bandwidth, and hence differences in performance are negligible, except for memory consumption, which is higher in Eiffel due to garbage collection (memory is deallocated only when necessary, thus the maximum memory usage is higher in operational conditions).

The test with \progname{libcurl} consisted in running all 583 tests from the standard testsuite mentioned before.
The total runtime is comparable in translated Eiffel and C.

The tests with \progname{libgmp} and \progname{libgsl} ran their respective standard testsuites.
The overall slow-down seems significant, but a closer look shows that the large majority of tests run in comparable time in C and Eiffel:
30\% of the \progname{libgmp} tests take up over 95\% of the running time; and 26\% of the \progname{libgsl} tests take up almost 99\% of the time.
The GCC torture tests incur only a moderate slow-down, concentrated in 3 tests that take 97\% of the time.
In all these experiments, the tests responsible for the conspicuous slow-down target operations that execute slightly slowlier in the translated Eiffel than in the native C (e.g., accessing a \cc{struct} field) and repeat it a huge number of times, so that the basic slow-down increases many-fold.
These bottlenecks are an issue only in a small fraction of the tests and could be removed manually in the translation.

The interactive applications (\progname{xeyes}, \progname{less}, \progname{links}, and \progname{vim}) run smoothly with good responsiveness, comparable to their original implementations.
The test for \progname{less} and \progname{vim} consisted in scrolling through large text files one line at a time (by holding ``arrow down'' in \progname{less}) or one page at a time (by holding ``page down'' in \progname{vim}).
Table~\ref{tab:eval2} shows the running times, which are the same in C and Eiffel; the screen refresh rate is also indistinguishable.

In all, the performance overhead in switching from C to Eiffel significantly varies with the program type but, even when it is significant, does not preclude the usability of the translated application or library in normal conditions---as opposed to the behavior in a few specific test cases.

\subsection{Usability and Maintainability} \label{sec:readability}
\noindent
A tool such as \nameSys, which provides automatic translation between languages, is applicable in different contexts within general software maintenance and evolution processes.
This section discusses some of these applications and how suitable \nameSys can be for each of them.

\titledpar{Reuse in clients.}
The first, most natural application is using \nameSys to automatically \emph{reuse} large C code-bases in Eiffel.
This is not merely a possibility, but something extremely valuable for Eiffel, whose user community is quite small compared to those of other mainstream languages such as C, Java, or C++.
Since we released \nameSys to the public as open-source, we have been receiving several requests from the community to produce Eiffel versions of C libraries whose functionalities are sorely missed in Eiffel, and whose native implementation would require a substantial effort to get to software of quality comparable to the widely tested and used C implementations.
This was the case, in particular, of \libname{libSDL\_mixer}, \libname{libmongoDB}, and \libname{libpcre}, whose automatic translations created using \nameSys were requested from the community and are now being used in Eiffel applications.
We are also aware of some Eiffel programmers directly trying to use \nameSys to translate useful C libraries and deploy them in their own software.

\begin{table}[!b]
\lstset{language=OOSC2Eiffel, escapechar=\!}
\begin{scriptsize}
\centering
\begin{tabular}{p{.455\textwidth} p{.48\textwidth}}
\begin{lstlisting}
class
   PRINT_SOURCE_SEHOME

inherit
   LIBCURL_CONSTANTS

create
   make

feature

   make
      local
         easy: P_EASY_STATIC
         handle: CE_POINTER [ANY]
         ret: NATURAL_32
      do
         create easy
         handle := easy.curl_easy_init
         easy.curl_easy_setopt ([handle, Curlopt_url,
           ce_string ("se.inf.ethz.ch")]).do_nothing
         ret := easy.curl_easy_perform (handle)
         easy.curl_easy_cleanup (handle)
      end
end
\end{lstlisting}
&
\begin{lstlisting}
class
   PRINT_SOURCE_SEHOME

inherit
   CURL_OPT_CONSTANTS

create
   make

feature

   make
      local
         easy: CURL_EASY_EXTERNALS
         handle: POINTER
         ret: INTEGER_32
      do
         create easy
         if easy.is_dynamic_library_exists then
            handle := easy.init
            easy.setopt_string (handle, Curlopt_url,
              "se.inf.ethz.ch")
            ret := easy.perform (handle)
            easy.cleanup (handle)
         else
            Io.error.put_string ("cURL $\mathit{not}$ found.%N")
         end
      end
end
\end{lstlisting}
\end{tabular}
\end{scriptsize}
\caption{Two Eiffel clients of cURL: using \libname{libcurl} translated with \nameSys (left) and using the library wrapper provided by EiffelStudio (right).}
\label{tab:curl-two-versions}
\end{table}

This form of reuse mainly entails writing Eiffel client code that accesses translated C components.
Supporting it requires tools that handle the full C language as used in practice, and that produce translated APIs understandable and usable with a programming style sufficiently close to what is the norm in Eiffel, without requiring in-depth understanding of the C conventions.
To give an idea of how \nameSys fares in this respect, consider writing a simple class \eif{PRINT_SOURCE_SEHOME}, which uses the API of cURL to retrieve and print the HTML code of the home page at \url{http://se.inf.ethz.ch}.
Table~\ref{tab:curl-two-versions} shows two versions of this client class.
The one on the left uses the translation of \progname{libcurl} automatically generated by \nameSys; the one on the right uses the wrapper of \progname{libcurl} part of the Eiffel standard libraries included with the EiffelStudio IDE, written by EiffelSoftware programmers.
The two solutions are quite similar in structure and style.
The version based on \nameSys still requires adhering to a couple of conventions that are a legacy of the translation from C: the arguments passed to routine \eif{curl_easy_setopt} must be listed in a tuple, and the routine call itself includes a \eif{do_nothing} which is needed whenever a function that returns a value is used as an instruction (Eiffel enforces separation between functions and procedures).
On the other hand, the ``native'' solution has a slightly more complex control structure, because it has to check that the dynamically linked cURL library is actually accessible at runtime; this is unnecessary with the implementation based on \nameSys, which does not depend on dynamically linked libraries since \progname{libcurl} is available translated to Eiffel.



When a C library undergoes maintenance, the introduced changes have the same impact on the C and on the Eiffel clients of the library.
In particular, if the changes to the library do not break client compatibility (that is, the API does not change), one should simply run \nameSys again on the new library version and replace it in the Eiffel projects that depends on it.
If the API changes, clients may have to change too, independent of the language they are written in.

\titledpar{Evolution of translated libraries.}
Once a program is translated from C to Eiffel, one may decide it has become part of the Eiffel ecosystem, and hence it will undergo maintenance and evolution as any other piece of Eiffel code.
In this scenario, \nameSys provides an immediately applicable solution to port C code to Eiffel, whereas the ensuing maintenance effort is distributed over an entire lifecycle and devoted to improve the automatic translation (for example, removing the application-dependent performance bottlenecks highlighted in Section~\ref{sec:performance}) and completely conforming it to the Eiffel style.

Such maintenance of translations is the easier the closer the generated code follows Eiffel conventions and, more generally, the object-oriented paradigm.
Providing a convincing empirical evaluation of the readability and maintainability of the code generated by \nameSys (not just from the perspective of writing client applications) is beyond the scope of the present work.
Notice, however, that \nameSys already follows numerous Eiffel conventions such as for the names of classes, types, and local variables, which might look verbose to hard-core C programmers but are de rigueur in Eiffel.
Follow-up work, which we briefly discuss in Section~\ref{sec:conc}, has targeted the object-oriented reengineering of \nameSys translations.
In all, while the translations produced by \nameSys still retain some ``C flavor'', we believe they are overall understandable and modifiable in Eiffel with reasonable effort.

\titledpar{Two-way propagation of changes.}
One more maintenance scenario occurs if one wants to be able to independently modify a C program and its translation to Eiffel, while still being able to propagate the changes produced in each to the counterpart.
For example, this scenario applies if a C library is being extended with new functionality, while its Eiffel translation produced by \nameSys undergoes refactoring to optimize it to the Eiffel environment.
This scenario is the most challenging of those discussed in this section; it poses problems similar to those of merging different development branches of the same project.
While merge conflicts are still a bane of collaborative development, modern version control systems (such as Git or Mercurial) have evolved to provide powerful support to ease the process of conflict reconciliation.
Thus, they could be very useful also in combination with automatic translators such as \nameSys to be able to integrate changes in C with other changes in Eiffel.

\subsection{Safety and Debuggability}\label{sec:bugs}
\noindent
Besides the obvious advantage of reusing the huge C code-base, translating C code to Eiffel using \nameSys leverages some high-level feature which may improve \emph{safety} and make \emph{debugging} easier in some conditions.

\titledpar{Uncontrolled format string} is a well-known vulnerability~\cite{UFS}
of C's \cc{printf} library function, which permits malicious clients to access data in the stack by supplying special format strings.
Consider for example the C program:

\begin{codesnip}
\begin{lstlisting}[language=CustomC]
     int main (int argc, char * argv[])
          { char *secret = "This is secret!"; if (argc > 1) printf(argv[1]); return 0; }
\end{lstlisting}
\end{codesnip}
If we call it with:
{\verb|./example "{Stack: %x%x%x%x%x%x} --> %s"|}, we get the output 
{\verb|{stack: 0b7|$[\ldots]$\verb|469} --> This is secret!|}, which reveals the local string \cc{secret}.
The safe way to achieve the intended behavior would be the instruction \cc{printf$\,$("\%s", argv[1])} instead of \cc{printf$\,$(argv[1])}, so that the input string is interpreted literally.

What is the behavior of code vulnerable to uncontrolled format
strings, when translated to Eiffel with \nameSys?  In simple usages of
\cc{printf} with just one argument as in the example, the translation
replaces calls to \cc{printf} with calls to Eiffel's
\eif{Io.put_string}, which
prints strings verbatim without interpreting them; therefore, the
translated code is not vulnerable in these cases. The replacement was
possible in 65\% of all the \cc{printf} calls in the programs of
Table~\ref{tab:eval}.  \nameSys translates
more complex usages of \cc{printf} (for example, with more than one
argument and no literal format string such as \cc{printf (argv[1],
  argv[2])}) into wrapped calls to the external \cc{printf} function,
and hence the vulnerability still exists.  However, it is
less extensive or more difficult to exploit in Eiffel: primitive types
(such as numeric types) are stored on the stack in Eiffel as they are
in C, but Eiffel's bulkier runtime typically stores them farther up
the stack, and hence longer and more complex format strings must be
supplied to reach the stack data (for instance, a variation of the example with \eif{secret} requires 386 \verb|%x|'s in the format string to reach local variables).
On the other hand, non-primitive types (such as strings
  and \cc{struct}s) are wrapped by Eiffel classes in \nameSys, which are stored in the heap, and hence
unreachable directly by reaching stack data.  In these cases, the
vulnerability vanishes in the Eiffel translation.

\titledpar{Debugging format strings.}
\nameSys also parses literal format strings passed to \cc{printf} and detects type mismatches between format specifiers and actual arguments.
This analysis, necessary when moving from C to a language with a stronger type system, helps debug incorrect and potentially unsafe usages of format strings.
Indeed, a mismatch detected while running the 145 \progname{libgmp} tests revealed a real error in the library's implementation of macro \cc{TESTS_REPS}:

\begin{codesnip}
\begin{lstlisting}[language=CustomC]
    char *envval, *end;    /* ... */
    long repfactor = strtol(envval, &end, 0);
    if(*end || repfactor $\leq$ 0) fprintf (stderr, "Invalid repfactor: %s.\n", repfactor);
\end{lstlisting}
\end{codesnip}
String \cc{envval} should have been passed to \cc{fprintf} instead of \cc{long repfactor}.
GCC with standard compilation options does not detect this error, which may produce garbage or even crash the program at runtime.
Interestingly, version 5.0.2 of \progname{libgmp} patches the code in the wrong way, changing the format specifier \cc{\%s} into \cc{\%ld}.
This is still incorrect because when \cc{envval} does not encode a valid ``repfactor'', the outcome of the conversion into \cc{long} is unpredictable.
Finally, notice that \nameSys may also report false positives, such as \cc{long v = "Hello!" ; printf("\%s", v)} which is acceptable (though probably not commendable) usage.

\titledpar{Out-of-bound error detection.}
C arrays translate to instances of class \eif{CE_ARRAY} (see Section~\ref{sec:types-type-constr}), which includes contracts that signal out-of-bound accesses to the array content.
Therefore, out-of-bound errors are much easier to detect in Eiffel applications using components translated with \nameSys.
Simply by translating and running the \progname{libgmp} testsuite, we found an off-by-one error causing out-of-bound access (our patch has been included in more recent versions of the library); the error does not manifest itself when running the original C version.
More generally, contracts help detect the precise location of array access errors.
Consider, for example:

\pagebreak
\begin{codesnip}
\begin{lstlisting}[language=CustomC,numbers=left,numberstyle=\tiny, numbersep=0.5pt, stepnumber=1]
             int * buf = (int *) malloc(sizeof (long long int) * 10);
             buf = buf - 10;  #\label{l:no-error}#
             buf = buf + 29;
             *buf = 'a';  buf++;
             *buf = 'b';  #\label{l:yes-error}#
\end{lstlisting}
\end{codesnip}
\cc{buf} is an array that stores 20 elements of type \cc{int} (which has half the size of \cc{long long int}).
The error is on line~\ref{l:yes-error}, when \cc{buf} points to position 20, out of the array bounds; line~\ref{l:no-error} is instead OK: \cc{buf} points to an invalid location, but it is not written to.
This program executes without errors in C; the Eiffel translation, instead, stops exactly at line~\ref{l:yes-error} and signals an out-of-bound access to \cc{buf}.

Array bound checking may be disabled, which is necessary in borderline situations where out-of-bound accesses do not crash because they assume a precise memory layout.
For example, \progname{links} and \progname{vim} use statements of the following form \linebreak \mbox{\cc{block *p = (block *) malloc(sizeof(struct block) + len)},}  with \cc{len > 0}, to allocate \cc{struct} datatypes of the form \cc{struct block \{ /* ... */ char b[1]; \}}.
In this case, \cc{p} points to a \cc{struct} with room for \cc{1 + len} characters in \cc{p$\rightarrow$b}; the instruction \cc{p$\rightarrow$b[len]=`i'} is then executed correctly in C, but the Eiffel translation assumes \cc{p$\rightarrow$b} has the statically declared size 1, hence it stops with an error.
Another borderline situation is with multi-dimensional arrays, such as \cc{double a[2][3]}.
An iteration over \cc{a}'s six elements with \cc{double *p = \&a[0][0]} translated to Eiffel fails to go past the third element, because it sees \cc{a[0][0]} as the first element of an array of length 3 (followed by another array of the same length).
A simple cast \cc{double * p = (double*)a} achieves the desired result without fooling the compiler, hence it works without errors also in translated code.
These situations are examples of unsafe programming more often than not.

\titledpar{More safety in Eiffel.}
Our experiments found another bug in \progname{libgmp}, where function \cc{gmp_sprintf_final} had three formal input arguments, but was only called with one actual through a function pointer.
Inspection suggests it is a copy-paste error of the other function \cc{gmp_sprintf_reps}.
The Eiffel version found the mismatch when calling the routine and reported a contract violation.
Easily finding such bugs demonstrates the positive side-effects of translating widely-used C programs into a tighter, higher-level language.





\subsection{Limitations} \label{sec:limitations}

The only significant limitations of the translation $\transl$ implemented in \nameSys in supporting C programs originate in the introduction of strong typing: programming practices that implicitly rely on a certain memory layout may not work in C applications translated to Eiffel.
Section~\ref{sec:bugs} mentioned some examples in the context of array manipulation (where, however, the checks on the Eiffel side can be disabled).
Another example is a function \cc{int trick (int a, int b)} that returns its second argument through a pointer to the stack, with the instructions \cc{int *p = \&a; return *(p+1)}.
\nameSys's translation assumes \cc{p} points to a single integer cell and cannot guarantee that \cc{b} is stored in the next cell.

Another limitation is the fact that \nameSys takes input from CIL, and hence it does not support legacy C such as K\&R C.
The support can, however, be implemented by directly extending the pre-processing CIL front-end.
Similarly, the GCC torture testsuite highlighted a few exotic GCC features currently unsupported by \nameSys (Section~\ref{sec:behavior}), which may be handled in future work.

Code using unrestricted \cc{goto}s poses the biggest hurdles to producing readable Eiffel code.
This is arguably unavoidable when translating to any programming language that does not have jumps.
The translation $\transl$ does, however, avoid the most complicated general translation scheme with the simpler control-flow breaking instructions \cc{continue}, \cc{break}, and \cc{return} instructions, whose translation is normally much more readable than when using unrestricted \cc{goto}s (see the example in Section~\ref{ex:longjmp}).



\section{Related Work}\label{sec:rw}

There are two main approaches to reuse source code written in a
``foreign'' language (e.g., C) in a different ``host''
language (e.g., Eiffel): define wrappers for the components
written in the foreign language; and translate the foreign
source code into functionally equivalent host code.
We discuss related work pursuing these approaches in Sections~\ref{rel:wrapping} and~\ref{rel:foreign}.
In Section~\ref{rel:OO}, we review the major solutions to automate object-oriented reengineering, which is a natural follow-up of automatic translation into object-oriented languages.

\subsection{Wrapping Foreign Code}\label{rel:wrapping}

Wrappers enable the reuse of foreign implementations through the API of bridge libraries.
This approach~(e.g., \cite{DietrichNackmanGracer89,deLuciaLuccaEtAl97,SerranoCarverOcaMontes02}) does not modify the foreign code, whose functionality is therefore not altered; moreover, the complete foreign language is supported.
On the other hand, the type of data that can be retrieved through the bridging API is often restricted to a simple subset common to the host and foreign language (e.g., primitive types).
\nameSys uses wrappers only to translate external functions and assembly code.

\subsection{Translating Foreign Code}\label{rel:foreign}

Industrial practices have long encompassed manual migrations of legacy code.
Some semi-automated tools exist that help translate code written in legacy programming languages such as old versions of COBOL~\cite{Terekhov00,Mossienko03}, Fortran-77~\cite{AcheeCarver97,SubramaniamByrne96}, and K\&R C~\cite{YehHarrisReubenstein95}.

Terekhov et al.~\cite{Terekhov00} review how automated language conversion is applied in industry; based on their experience, they conclude that ``creating 100\% automated conversion tools is neither possible, nor desirable''.
Our experience with \nameSys, however, suggests that such a conclusion has only relative validity: in the rich design space of automatic translators, there are scenarios where trading off some performance for complete automation is possible and desirable.
Rather than imposing an upfront heavyweight burden on developers in charge of migration, we suggest to start with an automatically translated version which is suboptimal but works out of the box, and then devote the manual programming effort to improving and adapting what is necessary---incrementally with an agile approach which also depends on the specific application domain and requirements.

Some translators focus on the adaptation of code into an extension
(superset) of the original language. Examples include the migration of
legacy code to object-oriented code, such as Cobol to
OO-Cobol~\cite{NewcombKotik95,Sneed92,WiggertsBosmaFielt97}, Ada to
Ada95~\cite{Sward04}, and C to
C++~\cite{KostasPrashant99,YingKostas01}.  Some of such efforts try to
go beyond the mere hosting of the original code, and introduce refactorings that take advantage of the object-oriented paradigm.
Most of these refactorings are, however, limited to restructuring modules into classes (see the focused discussion in Section~\ref{rel:OO}).
\nameSys follows a similar approach, but also takes advantage of some advanced features (such as contracts) to improve the reliability of translated code.

Ephedra~\cite{MartinMueller01} is a tool that translates legacy C to Java.
It first translates K\&R C to ANSI C; then, it maps data types and type casts; finally, it translate the C source code to Java.
Ephedra handles a significant subset of C, but cannot translate frequently used features such as unrestricted \cc{goto}s, external pre-compiled libraries, and inlined assembly
code.
A case study evaluating Ephedra~\cite{MartinMueller02} involved three small programs: the implementation of \cc{fprintf}, a monopoly game (about 3 KLOC), and two graph layout algorithms. The study reports that the source programs had to be manually adapted to be processable by Ephedra.
In contrast, \nameSys is completely automatic, and works with significantly larger programs.

Other tools (proprietary or open-source) to translate C (and C++) to Java or C\# include: C2J++~\cite{Tilevich97}, C2J~\cite{C2J}, and C++2C\# and C++2Java~\cite{Tangible}.
Table~\ref{tab:tools} shows a feature comparison among the currently available tools that translate C to an object-oriented language, showing:
\begin{itemize}
\item The \emph{target language}.
\item Whether the tool is \emph{completely automatic}, that is whether it generates translations that are ready for compilation. 
\item Whether the tool is \emph{available} for download and usable. In a couple of cases we could only find papers describing the tool but not a version of the implementation working on standard machines.
\item An (subjective to a certain extent) assessment of the \emph{readability} of the code produced.
In each case, we tried to evaluate if the translated code is sufficiently similar to the C source to be readily understandable by a programmer familiar with the latter.
We judged \progname{C2J}'s readability negatively because the tool puts data into a single global array to support pointer arithmetic. This is quite detrimental to readability and also circumvents type checking in the Java translation.
\item Whether the tool supports unrestricted calls to \emph{external libraries}, unrestricted \emph{pointer arithmetic}, unrestricted \cc{goto}s, and inlined \emph{assembly code}.
\end{itemize}
\noindent
The table demonstrates that \nameSys is arguably the first completely automatic tool that handles the complete C language as used in practice.

\begin{table}
\caption{Tools translating C to O-O languages.}\label{tab:tools}
\centering
\begin{tabular}{lcccccccc}
\toprule
  &
  \begin{sideways}\parbox{1.5cm}{target\\language}\end{sideways} &
  \begin{sideways}\parbox{1.5cm}{completely\\automatic}\end{sideways} &
  \begin{sideways}available\end{sideways} &
  \begin{sideways}\parbox{1.5cm}{readability}\end{sideways} &
  \begin{sideways}\parbox{1.5cm}{external\\libraries}\end{sideways} &
  \begin{sideways}\parbox{1.5cm}{pointer\\arithmetic}\end{sideways} &
  \begin{sideways}\cc{goto}s\end{sideways} &
  \begin{sideways}\parbox{1.5cm}{inlined\\assembly}\end{sideways}\\
\midrule
\progname{Ephedra}      & Java   & no  & no  & $+$ & no  & no  & no  & no  \\
\progname{Convert2Java} & Java   & no  & no  & $+$ & no  & no  & no  & no  \\
\progname{C2J++}        & Java   & no  & no  & $+$ & no  & no  & no  & no  \\
\progname{C2J}          & Java   & no  & yes & $-$ & no  & yes & no  & no  \\
\progname{C++2Java}     & Java   & no  & yes & $+$ & no  & no  & no  & no  \\
\progname{C++2C\#}      & C\#    & no  & yes & $+$ & no  & no  & no  & no  \\
\hline
\progname{C2Eif}        & Eiffel & yes & yes & $+$ & yes & yes & yes & yes \\
\bottomrule
\end{tabular}
\end{table}

In previous work, we developed J2Eif, an automatic source-to-source
translator from Java to Eiffel~\cite{TOFN11-TOOLS11}; translating
between two object-oriented languages does not pose some of the
formidable problems of bridging wildly different abstraction levels,
which \nameSys had to deal with.

TXL~\cite{Cordy06} is an expressive programming language designed to support source code analysis and transformation.
It has been used to implement several language translation frameworks such as from Java to TCL and to Python.
We could have used TXL to implement \nameSys; using regular Eiffel, however, allowed us to be independent of third-party closed-source tools, and to retain complete control over the implemented functionalities.

\titledpar{Safer C.}
Many techniques exist aimed at ameliorating the safety of existing C code; for example, detection of format string vulnerability~\cite{FGuard}, out-of-bound array accesses and other memory errors~\cite{Oiwa2009,NethercoteS07}, or type errors~\cite{Necula02}.
\nameSys has a different scope, as it offers improved safety and debuggability as \emph{side-benefits} of automatically porting C programs to Eiffel.
This shares a little similarity with Ellison and Rosu's formal executable semantics of C~\cite{EllisonRosu12}, which also finds errors in C program as a ``side effect'' of a rigorous translation.

\subsection{Object-Oriented Reengineering}\label{rel:OO}
After code written in a procedural language has been migrated to an object-oriented environment, it is natural to reengineer it to conform to the object-oriented design style, taking full advantage of features such as inheritance; this is the goal of object-oriented reengineering.
Object-oriented reengineering is beyond the scope of this paper; and we tackled it in follow-up work based on \nameSys which we mention in Section~\ref{sec:conc}.
Nonetheless, it is still useful to compare existing approaches to reengineering with \nameSys,  solely based on features such as degree of automation and tool support; see our follow-up work~\cite{ECOOP-paper} for a discussion focused on the object-oriented reengineering techniques.

\begin{table}
\caption{Comparison of approaches to O-O reengineering.}\label{tab:rw}
\centering
\begin{tabular}{lccccccccccccc}
\toprule
  &~~~&
  \begin{sideways}source--target\end{sideways} &~~~&
  \begin{sideways}{tool support}\end{sideways} &~~~&
  \begin{sideways}\parbox{1.9cm}{completely\\automatic}\end{sideways} &~~~&
  \begin{sideways}full language\end{sideways} &~~~&
  \begin{sideways}evaluated\end{sideways}  \\
\midrule
Gall~\cite{GK}                &&  methodology  &&  no            &&  no   &&  --   &&  yes      \\
Jacobson~\cite{Ivar}          &&  methodology  &&  no            &&  no   &&  --   &&  yes      \\
\hline
Livadas~\cite{LJ}             &&  C--C++       &&  yes           &&  no   &&  no   &&  no       \\
Kontogiannis~\cite{KP}        &&  C--C++       &&  yes           &&  no   &&  ?    &&  10KL     \\
Frakes~\cite{Frakes08}        &&  C--C++       &&  yes           &&  no   &&  no   &&  2KL      \\
Fanta~\cite{Fanta98}          &&  C++--C++     &&  yes           &&  no   &&  no   &&  120KL    \\
\hline
Newcomb~\cite{NK}             &&  Cobol--OOSM  &&  yes           &&  yes  &&  no   &&  168KL    \\
Mossienko~\cite{Mossienko03}  &&  Cobol--Java  &&  yes           &&  no   &&  no   &&  25KL     \\
Sneed~\cite{Sneed10}          &&  Cobol--Java  &&  yes           &&  yes  &&  no   &&  200KL    \\
Sneed~\cite{Sneed2011}        &&  PL/I--Java   &&  yes           &&  yes  &&  no   &&  10KL     \\
\hline
\nameSys                      &&  C--Eiffel    &&  \textsc{yes}  &&  yes  &&  yes  &&  932KL    \\
\bottomrule
\end{tabular}
\end{table}

Table~\ref{tab:rw} summarizes some features of the main approaches to object-oriented reengineering of procedural code:

\begin{itemize}
\item The \textit{source} and the \emph{target} languages (or if it is a generic methodology).
\item Whether \emph{tool support} was developed, that is whether there exists a tool or the paper explicitly mentions the implementation of a tool. A \textsc{yes} in small caps denotes the only currently publicly available tool, namely \nameSys.
\item Whether the approach is \emph{completely automatic}, that is does not require any user input other than providing a source procedural program.
\item Whether the approach supports the \textit{full} source \textit{language} (as used in practice) or only a subset thereof.
\item Whether the approach has been \emph{evaluated}, that is whether the paper mentions evidence, such as a case study, that the approach was tried on real programs. If available, the table indicates the size of the programs used in the evaluation.

\end{itemize}

Newcomb's~\cite{NK} and Sneed's~\cite{Sneed10} are the only automatic tools which have been evaluated on programs of significant size.
Newcomb's tool~\cite{NK}, however, produces hierarchical object-oriented state machine models (OOSM); the mapping from OOSM to an object-oriented language is out of the scope of the work.
Sneed's tool~\cite{Sneed10} translates Cobol to Java; the paper reports that manual corrections of the automatically generated code were needed to get to a correct translation.
While these corrections have successively been incorporated as extensions of the tool, the
full Cobol language remains unsupported, according to the paper.


\section{Conclusions and Future Work}\label{sec:conc}

This paper presented the complete translation of C applications into Eiffel, and its implementation into the freely available automatic tool \nameSys.
\nameSys supports the complete C language, including unrestricted pointer arithmetic and pre-compiled libraries.
Experiments in the paper showed that \nameSys correctly translates complete applications and libraries of significant size, and
takes advantage of some of Eiffel's advanced features to produce code of good quality.

\titledpar{Future work.}
Future work will improve the readability and maintainability of the generated code.
CIL, in particular, optimizes the code for program analysis, which is sometimes detrimental to readability of the Eiffel code generated by \nameSys.
For example, CIL does not preserve comments, which are therefore lost in translation.
We will also optimize the helper classes to improve on the few performance bottlenecks mentioned in Section~\ref{sec:performance}.

A major follow-up to the work described in this paper is the object-oriented reengineering of C code translated to Eiffel.
In recent work~\cite{ECOOP-paper}, we have developed an automatic reengineering technique, and implemented it atop the translation produced by \nameSys and described in the present paper.
The technique encapsulates functions and type definitions into classes that achieve low coupling and high cohesion, and introduces inheritance and contracts when possible.
Other future work still remains in the direction of reengineering, such as in automatically replacing C data structure implementations (e.g., hash tables) with their Eiffel equivalents.



\begin{thebibliography}{10}

\bibitem{AcheeCarver97}
B.~L. Achee and Doris~L Carver.
\newblock Creating object-oriented designs from legacy {FORTRAN} code.
\newblock {\em Journal of Systems and Software}, 39(2):179--194, 1997.

\bibitem{Cordy06}
James~R. Cordy.
\newblock The {TXL} source transformation language.
\newblock {\em Sci. Comput. Program.}, 61(3):190--210, 2006.

\bibitem{FGuard}
Crispin Cowan.
\newblock {FormatGuard}: Automatic protection from \texttt{printf} format
  string vulnerabilities.
\newblock In {\em Proceedings of the 10th USENIX Security Symposium}, 2001.

\bibitem{UFS}
{CWE-134}.
\newblock Uncontrolled format string.
\newblock \url{http://cwe.mitre.org/data/definitions/134.html}.

\bibitem{deLuciaLuccaEtAl97}
Andrea de~Lucia, Giuseppe A.~Di Lucca, Anna~Rita Fasolino, Patrizia Guerra, and
  Silvia Petruzzelli.
\newblock Migrating legacy systems towards object-oriented platforms.
\newblock {\em Proc. of ICSM}, pages 122--129, 1997.

\bibitem{DietrichNackmanGracer89}
W.~C. Dietrich, Jr., L.~R. Nackman, and F.~Gracer.
\newblock Saving legacy with objects.
\newblock {\em SIGPLAN Not.}, 24(10):77--83, 1989.

\bibitem{EllisonRosu12}
Chucky Ellison and Grigore Rosu.
\newblock An executable formal semantics of {C} with applications.
\newblock In {\em POPL}, pages 533--544, 2012.

\bibitem{Fanta98}
R.~Fanta and V.~Rajlich.
\newblock Reengineering object-oriented code.
\newblock In {\em International Conference on Software Maintenance, 1998.
  Proceedings}, pages 238--246, 1998.

\bibitem{libffi}
FFI.
\newblock A portable foreign function interface library.
\newblock \url{http://sources.redhat.com/libffi/}, 2011.

\bibitem{Frakes08}
William Frakes, Gregory Kulczycki, and Natasha Moodliar.
\newblock An empirical comparison of methods for reengineering procedural
  software systems to object-oriented systems.
\newblock In {\em ICSR}, volume 5030 of {\em LNCS}, pages 376--389, 2008.

\bibitem{GK}
H.~Gall and R.~Klosch.
\newblock Finding objects in procedural programs: an alternative approach.
\newblock In {\em WCRE}, pages 208--216. IEEE, 1995.

\bibitem{GJ-PL}
Carlo Ghezzi and Mehdi Jazayeri.
\newblock {\em Programming Language Concepts}.
\newblock Wiley, 3rd edition, 1998.

\bibitem{Harel-folk}
David Harel.
\newblock On folk theorems.
\newblock {\em Commun. ACM}, 23(7):379--389, 1980.

\bibitem{Ivar}
Ivar Jacobson and Fredrik Lindstr\"{o}m.
\newblock Reengineering of old systems to an object-oriented architecture.
\newblock In {\em OOPSLA}, pages 340--350. ACM, 1991.

\bibitem{KostasPrashant99}
Kostas Kontogiannis and Prashant Patil.
\newblock Evidence driven object identification in procedural code.
\newblock In {\em STEP}, pages 12--21, 1999.

\bibitem{KP}
Kostas Kontogiannis and Prashant Patil.
\newblock Evidence driven object identification in procedural code.
\newblock In {\em STEP}, pages 12--21. IEEE, 1999.

\bibitem{LJ}
Panos~E. Livadas and Theodore Johnson.
\newblock A new approach to finding objects in programs.
\newblock {\em Journal of Software Maintenance}, 6(5):249--260, 1994.

\bibitem{MartinMueller01}
Johannes Martin and Hausi~A. M\"uller.
\newblock Strategies for migration from {C} to {J}ava.
\newblock In {\em CSMR}, pages 200--210. IEEE Computer Society, 2001.

\bibitem{MartinMueller02}
Johannes Martin and Hausi~A. M\"uller.
\newblock C to {J}ava migration experiences.
\newblock In {\em CSMR}, pages 143--153. IEEE Computer Society, 2002.

\bibitem{OOSC2}
Bertrand Meyer.
\newblock {\em {Object-Oriented Software Construction}}.
\newblock Prentice Hall, 2nd edition, 1997.

\bibitem{Mossienko03}
M.~Mossienko.
\newblock Automated {C}obol to {J}ava recycling.
\newblock In {\em CSMR}, pages 40--50. IEEE, 2003.

\bibitem{NeculaMcPeakRahulWeimer}
George~C. Necula, Scott McPeak, S.P. Rahul, and Westley Weimer.
\newblock {CIL}: {I}ntermediate {L}anguage and {T}ools for {A}nalysis and
  {T}ransformation of {C} {P}rograms.
\newblock In {\em Conference on Compilier Construction}, pages 213--228, 2002.

\bibitem{Necula02}
George~C. Necula, Scott McPeak, and Westley Weimer.
\newblock {CCured}: type-safe retrofitting of legacy code.
\newblock In {\em POPL}, pages 128--139, 2002.

\bibitem{NethercoteS07}
Nicholas Nethercote and Julian Seward.
\newblock Valgrind: a framework for heavyweight dynamic binary instrumentation.
\newblock In {\em PLDI}, pages 89--100, 2007.

\bibitem{NewcombKotik95}
P.~Newcomb and G.~Kotik.
\newblock Reengineering procedural into object-oriented systems.
\newblock In {\em WCRE}, pages 237--249, 1995.

\bibitem{NK}
P.~Newcomb and G.~Kotik.
\newblock Reengineering procedural into object-oriented systems.
\newblock In {\em WCRE}, pages 237--249. IEEE, 1995.

\bibitem{C2J}
Novosoft.
\newblock {C2J}: a {C} to {J}ava translator.
\newblock \url{http://www.novosoft-us.com/solutions/product_c2j.shtml}, 2001.

\bibitem{Oiwa2009}
Yutaka Oiwa.
\newblock Implementation of the memory-safe full {ANSI-C} compiler.
\newblock In {\em Proceedings of the 2009 ACM SIGPLAN conference on Programming
  language design and implementation}, PLDI '09, pages 259--269, New York, NY,
  USA, 2009. ACM.

\bibitem{C-history}
Dennis Ritchie.
\newblock The development of the {C} language.
\newblock In {\em History of Programming Languages Conference (HOPL-II)
  Preprints}, pages 201--208, 1993.

\bibitem{SerranoCarverOcaMontes02}
Miguel~A. Serrano, Doris~L. Carver, and Carlos~Montes de~Oca.
\newblock Reengineering legacy systems for distributed environments.
\newblock {\em J. Syst. Softw.}, 64(1):37--55, 2002.

\bibitem{Sneed2011}
H.~Sneed.
\newblock Migrating {PL/I} code to {J}ava.
\newblock In {\em CSMR}, pages 287--296. IEEE, 2011.

\bibitem{Sneed92}
H.M. Sneed.
\newblock Migration of procedurally oriented {C}obol programs in an
  object-oriented architecture.
\newblock In {\em Software Maintenance}, pages 105--116, 1992.

\bibitem{Sneed10}
H.M. Sneed.
\newblock Migrating from {COBOL} to {J}ava.
\newblock In {\em ICSM}, pages 1--7. IEEE, 2010.

\bibitem{SubramaniamByrne96}
Gokul~V. Subramaniam and Eric~J. Byrne.
\newblock Deriving an object model from legacy {F}ortran code.
\newblock {\em ICSM}, pages 3--12, 1996.

\bibitem{Sward04}
Ricky Sward.
\newblock Extracting ada 95 objects from legacy ada programs.
\newblock In Albert Llamos\'i and Alfred Strohmeier, editors, {\em Reliable
  Software Technologies - Ada-Europe 2004}, volume 3063 of {\em Lecture Notes
  in Computer Science}, pages 65--77. Springer, 2004.

\bibitem{Tangible}
{Tangible Software Solutions}.
\newblock {C++} to {C\#} and {C++} to {J}ava.
\newblock \url{http://www.tangiblesoftwaresolutions.com/}.

\bibitem{Terekhov00}
Andrey~A. Terekhov and Chris Verhoef.
\newblock The realities of language conversions.
\newblock {\em IEEE Software}, 17(6):111--124, 2000.

\bibitem{popularity}
{The Language Popularity Index}.
\newblock \url{http://lang-index.sourceforge.net}, 2011.

\bibitem{Tilevich97}
Eli Tilevich.
\newblock Translating {C++} to {J}ava.
\newblock In {\em First German Java Developers' Conference Journal, Sun
  Microsystems Press}. IEEE Computer Society, 1997.

\bibitem{C2Eif-demo}
Marco Trudel, Carlo~A. Furia, and Martin Nordio.
\newblock Automatic {C} to {O-O} translation with {C2Eiffel}.
\newblock In Rocco Oliveto, Denys Poshyvanyk, James Cordy, and Thomas Dean,
  editors, {\em Proceedings of the 19th Working Conference on Reverse
  Engineering (WCRE'12)}, pages 508--509. IEEE Computer Society, October 2012.
\newblock Tool demonstration paper.

\bibitem{ECOOP-paper}
Marco Trudel, Carlo~A. Furia, Martin Nordio, and Bertrand Meyer.
\newblock Really automatic scalable object-oriented reengineering.
\newblock In Giuseppe Castagna, editor, {\em Proceedings of the 27th European
  Conference on Object-Oriented Programming (ECOOP)}, volume 7920 of {\em
  Lecture Notes in Computer Science}, pages 477--501. Springer, July 2013.

\bibitem{TFNMO12-WCRE12}
Marco Trudel, Carlo~A. Furia, Martin Nordio, Bertrand Meyer, and Manuel Oriol.
\newblock {C} to {O-O} translation: Beyond the easy stuff.
\newblock In Rocco Oliveto, Denys Poshyvanyk, James Cordy, and Thomas Dean,
  editors, {\em Proceedings of the 19th Working Conference on Reverse
  Engineering (WCRE'12)}, pages 19--28. IEEE Computer Society, October 2012.

\bibitem{TOFN11-TOOLS11}
Marco Trudel, Manuel Oriol, Carlo~A. Furia, and Martin Nordio.
\newblock Automated translation of {J}ava source code to {E}iffel.
\newblock In {\em TOOLS Europe}, volume 6705 of {\em LNCS}, pages 20--35, 2011.

\bibitem{TFNM11-SEFM11}
Julian Tschannen, Carlo~A. Furia, Martin Nordio, and Bertrand Meyer.
\newblock Usable verification of object-oriented programs by combining static
  and dynamic techniques.
\newblock In {\em SEFM}, volume 7041 of {\em LNCS}, pages 382--398, 2011.

\bibitem{WiggertsBosmaFielt97}
Theo Wiggerts, Hans Bosma, and Erwin Fielt.
\newblock Scenarios for the identification of objects in legacy systems.
\newblock In {\em WCRE}, pages 24--32, 1997.

\bibitem{YehHarrisReubenstein95}
A.~Yeh, D.~Harris, and H.~Reubenstein.
\newblock Recovering abstract data types and object instances from a
  conventional procedural language.
\newblock In {\em WCRE}, pages 227--236, 1995.

\bibitem{YingKostas01}
Ying Zou and Kostas Kontogiannis.
\newblock A framework for migrating procedural code to object-oriented
  platforms.
\newblock In {\em APSEC}, pages 390--399, 2001.

\end{thebibliography}

\end{document}